\documentstyle[fleqn,psfig]{l-aa}
\newcommand{\ms}{ M$_{\odot}$ }

\newcommand{\mss}{M$_{\odot}$}
\newenvironment{eqnum}[1]{\begin{center} \vspace{-1.2cm} \hfill (#1) 
\vspace{0.1cm}}{\end{center}}

\newcommand{\eg}{\rm e.g.\ }
\newcommand{\etal}{\rm et al. }
\newcommand{\reflistset}{\setlength{\leftmargin}{1.0cm}
        \setlength{\itemindent}{-0.6cm}\setlength{\topsep}{0.0cm}
        \setlength{\itemsep}{-0.093cm}}
\newcommand{\ie}{\rm i.e.\/ }

\newcommand{\cf}{cf.\/ }

\begin{document}

\thesaurus{08(09.01.1; 10.01.1; 10.05.1; 10.19.1; 11.09.4)}

\title{ Inhomogeneous chemical evolution of the Galactic disk:
evidence for sequential stellar enrichment?}
\author{L.B. ~van den Hoek\inst{1} and T. ~de Jong\inst{1,2} }
\institute{Astronomical Institute 'Anton Pannekoek',
   Kruislaan 403, NL 1098 SJ Amsterdam, 
   The Netherlands
\and
   Space Research Organisation of the 
   Netherlands, Sorbonnelaan 2, NL 3584 CA Utrecht, The Netherlands}
\date{Accepted 19/6/96}

\offprints{L.B. van den Hoek (bobby@astro.uva.nl)}

\maketitle
\markboth{L.B. van den Hoek \& T. de Jong: Inhomogeneous chemical evolution 
of the Galactic disk}{}

\psnoisy
%\psdraft
%\psrotatefirst
%\setcounter{page}{2}

\begin{abstract} 

We investigate the origin of the abundance variations observed 
among similarly aged F and G dwarfs in the local Galactic disk. 
We argue that orbital diffusion of stars in combination with radial abundance 
gradients is probably insufficient to explain these variations. 

We show that episodic and local infall of metal-deficient gas can 
provide an adequate explanation for iron and oxygen variations as 
large as $\Delta$[M/H] $\sim$0.6 dex among stars formed at a given age in the 
solar neighbourhood (SNBH). However, such models appear inconsistent 
with the observations because they: 1) result in current disk ISM 
abundances that are too high compared to the observations, 2) predict 
stellar abundance variations to increase with the lifetime of the 
disk, and 3) do not show substantial scatter in the [Fe/H] vs. [O/H] relation.
Notwithstanding, our results do
suggest that metal-deficient gas infall plays an important role in regulating 
the chemical evolution of the Galactic disk.

We demonstrate that sequential enrichment by successive stellar generations 
within individual gas clouds can account for substantial abundance variations 
as well. However, such models are inconsistent with the observations because 
they: 1) are unable to account for the full magnitude of the observed 
variations, in particular for [Fe/H], 2) predict stellar abundance 
variations to decrease with the lifetime of the disk, and 3) result in 
current abundances far below the typical abundances observed in the local 
disk ISM.

We present arguments in support of {\em combined} infall of metal-deficient 
gas and sequential enrichment by successive stellar generations in the local 
Galactic disk ISM. We show that galactic chemical evolution models 
which take into account these processes simultaneously are consistent with 
both the observed abundance variations among similarly aged F and G dwarfs in 
the SNBH {\em and} the abundances observed in the local disk ISM.
For reasonable choices of parameters, these models can reproduce 
$\Delta$[M/H] for individual elements M = C, O, Fe, Mg, Al, and Si as well as 
the scatter observed in abundance-abundance relations like [O/Fe]. 
For the same models, the contribution of sequential stellar enrichment to the 
magnitude of the observed abundance variations can be as large as 
$\sim$50\%.

We discuss the impact of sequential stellar enrichment
and episodic infall of metal-deficient gas on the inhomogeneous chemical 
evolution of the Galactic disk.

\end{abstract}

\section{Introduction}

The chemical enrichment of the interstellar medium (ISM) by 
successive generations of stars is a key issue in understanding 
the chemical evolution of galaxies in general, and the formation history 
and abundance distributions of the stellar populations in our Galaxy in 
particular.
Observational studies related to the heavy element enrichment of the local 
Galactic disk have long shown that stars of similar age exhibit large 
abundance variations (\eg Mayor 
1976; Twarog 1980a; Twarog \& Wheeler 1982; 
Carlberg \etal 1985; Gilmore 1989; Klochkova \etal 1989; Schuster \& 
Nissen 1989; Meusinger \etal 1991).
Recently, Edvardsson \etal (1993a) presented accurate abundance data 
for nearly 200 F and G main-sequence dwarfs in the solar neighbourhood (SNBH).
Their spectroscopic data, analysed with up-to-date input physics, confirms  
abundance variations as large as $\sim$0.6 dex in 
$\Delta$[M/H] (where M=Fe,O,Mg,Al,Si) among similarly aged stars.
In contrast to previous understanding, these variations are much in excess 
of experimental uncertainties
and demonstrate that the abundance spread {\em for stars born at 
roughly the same galactocentric distance} is similar in magnitude to the 
overall increase in metallicity during the lifetime of the disk. 

Additional support for the existence of large abundance inhomogeneities in 
the Galactic disk has been provided by studies of stars in open clusters 
(\eg Nissen 1988; Boesgaard 1989; Lambert 1989; Garci\'{a}-Lopez \etal 1993; 
Friel \& Janes 1993: Carraro \& Chiosi 1994) and B stars in star forming 
regions in the SNBH 
(\eg Gies \& Lambert 1992; Cunha \& Lambert 1992).
These studies show that the concept of a well-defined tight age-metallicity 
relation (AMR) for the Galactic disk ISM is unfounded (Edmunds 1993) and that 
the chemical enrichment of the disk has been inhomogeneous 
on time scales as short as $\sim$10$^{8}-10^{9}$ yr.
Similar studies of objects in the Magellanic clouds 
(\eg Cohen 1982; Da Costa 1991; Olsewski \etal 1991) and 
dwarf galaxies (Pilyugin 1992; Kunth \etal 1994; Thuan \etal 1995) suggest 
that inhomogeneous chemical evolution is a common phenomenon in 
nearby galaxies as well.

The origin of the abundance variations observed in 
the local Galactic disk is investigated in this 
paper. Clearly, large abundance variations in the ISM on time scales 
at least an order of magnitude shorter than the lifetime of the disk cannot 
be reproduced by simple galactic evolution models incorporating 
monotonously increasing age-metallicity relations (AMR).
In the past few years, various ideas have been put forward as possible 
explanations for the intrinsic abundance variations among similarly aged 
stars:
\begin{list}{$\bullet$}{\leftmargin 0.65cm}
\item stellar orbital diffusion in combination with radial abundance gradients 
in the Galactic disk (\eg Francois \& Matteucci 1993; Wielen, Fuchs, \& 
Dettbarn 1996);
\item sequential enrichment by successive stellar generations (\eg Edmunds 1975;
Olive \& Schramm 1982; Gilmore 1989; Gilmore \& Wyse 1991; Cunha \& Lambert 
1992; Roy and Kunth 1995); 
\item local infall of metal-poor gas (\eg Edvardsson \etal 1993a; Roy \& 
Kunth 1995; Pilyugin \& Edmunds 1995a);
\item cloud motions in the ISM (Bateman \& Larson 1993);
\item inefficient mixing in the disk ISM: isolated chemical evolution of 
individual parcels of interstellar gas during considerable fractions of the 
lifetime of the disk (Lennon \etal 1990; Wilmes \& K\"{o}ppen 1995); 
\item major galaxy merger events resulting in multiple stellar populations 
in the Galactic disk (Strobel 1991; Pilyugin \& Edmunds 1995b);
\item chemical fractionization processes such as grain formation 
(\eg Henning \& G\"{u}rtler 1986) and/or element 
diffusion (\eg Bahcall \& Pinsonneault 1995) so that measured abundances do 
not reflect initial stellar abundances. 
\end{list}

As discussed by Edvardsson \etal (1993a) stellar orbital 
diffusion is probably inadequate as main explanation for the observed 
abundance variations. However, recently Wielen \etal (1996) claimed that 
stars can be born at galactocentric distances very different from those 
derived using their present-day orbits. In this case, a major fraction of 
the observed abundance scatter could be due to stellar orbital diffusion.
Since it appears unlikely that diffusion can explain the observed 
abundance variations for all stars in the Edvardsson \etal sample (see below) 
as well as those observed among young stars (\eg present in star forming 
regions) other processes are probably important as well.

Based on the assumption of short mixing 
time scales of $\sim 10^{7}$ yr in the local disk ISM, 
we argue that the underlying physical mechanisms causing the observed 
abundance inhomogeneities and those initiating star formation in the disk ISM 
are the same. No observational support exists for chemical 
fractionization in low mass F and G main-sequence stars. Therefore, the 
scatter in stellar metallicities probably reflects the original 
inhomogeneities in the interstellar gas (\eg Gilmore 1989).
We here restrict ourselves mainly to the processes of 
sequential stellar enrichment and episodic infall of gas onto the Galactic 
disk as possible explanation for the observed stellar abundance variations.
Since these processes are observed to operate simultaneously in the SNBH 
(see below), it is important to investigate their combined effect on the 
chemical evolution of the Galactic disk.

The process of star formation initiated by stars formed during a preceding 
star formation event nearby, is known as sequential star formation. 
Support for sequential star formation in the SNBH is provided 
by observations of spatially separated subgroups of OB stars that appear 
aligned in a sequence of ages in many OB associations (\eg Blaauw 1991) and 
by obervations of stars forming at the interfaces of H{\sc ii} regions and 
their surrounding molecular clouds (\eg Genzel \& Stutzki 1989; Pismis 1990; 
Goldsmith 1995). Sequential star formation may be induced by the blast waves of 
nearby supernova explosions compressing the ambient ISM (\eg Ogelman \& 
Maran 1976) and/or by propagating ionization and shock fronts from an OB 
association causing the gravitational collapse of a nearby molecular cloud
(\eg Elmegreen \& Lada 1977). In either case, efficient self-enrichment 
through mixing of enriched material by successive generations of massive stars 
is expected.

On the other hand, stellar abundance variations can be attributed to
infall of relatively unprocessed gas and star formation within the 
accreted material before efficient mixing wipes out any local chemical 
inhomogeneities (\eg Edvardsson \etal 1993a; Pilyugin \& Edmunds 1995a).
Observational support for star formation in the SNBH initiated by 
infall of high velocity clouds, has
recently been presented by L\'{e}pine \& Duvert (1994).
These authors claim that episodic gas infall is a dominant process in the 
local disk ISM and is associated with all prominent star forming molecular 
clouds seen near the Sun (see Sect. 5.2).
Furthermore, ongoing gas infall has been emphasized by models of dissipative 
protogalactic collapse (Larson 1969, 1976) and on the basis of time
scale arguments of gas consumption in the local disk (Larson \etal 1980; 
Kennicutt 1983).

In this paper, we present a chemical evolution model for a star forming gas 
cloud 
which incorporates stellar enrichment and mixing processes (including infall) 
and which allows for temporal and/or spatial inhomogeneities in the ISM.
This study differs from previous work (\eg Pilyugin \& Edmunds 1995a) in 
that we investigate in detail the combined effect of metal deficient gas 
infall and sequential stellar enrichment by successive stellar generations 
on the chemical evolution of multiple gas clouds in the Galactic disk.
In particular, each gas cloud is allowed to follow its individual star 
formation, mixing, and infall history, as is suggested by the observations.
With this model, 
we fit the stellar abundance variations and current local ISM abundances of 
C, O, Fe, Mg, Al, and Si observed in the SNBH, the present gas-to-total 
mass-ratio, and actual star formation and supernova rates. We note that 
previous investigations were restricted to abundance variations in oxygen 
and iron only.

We will show that models taking into account the above processes simultaneously 
are in good agreement with the observations and provide an adequate
explanation for the stellar abundance variations with respect to the 
mean abundances observed in the local disk.
Furthermore, we will argue that the contribution of sequential stellar 
enrichment to the magnitude of the observed stellar abundance variations 
can be as large as $\sim$50\%, \ie much larger than suggested by previous 
investigations (\eg Pilyugin \& Edmunds 1995a; Wilmes and K\"{o}ppen 1995).
Corresponding theoretical age and abundance-distributions related to 
the G-dwarf problem will be discussed in a separate paper.

The paper is organized as follows. 
In Sect. 2, we briefly review observations related to the inhomogeneous
heavy element enrichment of the local Galactic disk ISM.
In Sect. 3, we describe characteristics of the inhomogeneous 
chemical evolution model proposed for the Galactic disk 
(model equations and details are given in the Appendix to the electronic 
version of this paper). 
In Sect. 4, we present model results 
for episodic infall of metal-deficient gas and 
sequential stellar enrichment, and examine which of these mechanisms can 
account satisfactorily for the observations.
In Sect. 5, we discuss these results in the more general context of the 
chemical evolution of the Galactic disk and adduce both observational 
arguments in support of sequential star formation and metal deficient gas 
infall in the local disk ISM.

\section{Inhomogeneous chemical evolution of the local Galactic disk: 
observations}

\subsection{Main-sequence F and G dwarfs}

We concentrate on the abundance data of nearly 200 main-sequence field F and G 
dwarfs with actual distances $\la$70 pc from the Sun as recently presented 
by Edvardsson \etal (1993a; hereafter EDV). This sample provides the largest 
sample of stars available to date for studies related to the chemical 
evolution of the local disk. Fig. 1 displays all F and G dwarfs for
which both [O/H] and [Fe/H] abundance-ratios have been determined by EDV. 

Large abundance variations of $\sim$0.9 dex in [Fe/H] and 
$\sim$0.7 dex in [O/H] among stars of a given age are seen to be present 
(abundance variations in \eg Mg, Al, and Si resemble those in [Fe/H]).
At intermediate stellar ages, these
variations are no doubt significant since typical observational 
errors are $\sim$0.1 dex both in 
[M/H] and log(Age) (see EDV). At ages in excess of $\sim$15 Gyr and less than 
$\sim$2 Gyr, the data probably are undersampled (see EDV). Note that
the sample is biased against old, high-metallicity stars through the minimum 
$T_{\rm eff}$ limit assumed by EDV. 

The observed spread in [Fe/H] is tightly correlated with that in 
[O/H]. This suggests that different nucleo-synthesis sites, 
which contribute different elements to the initial abundances in stars, mix 
their products together well. Furthermore, this suggests that stellar abundance 
variations for different elements are due to the same process.
Current observations 
support the idea that the magnitude of the stellar abundance variations has 
remained constant over the lifetime of the disk (see also Mayor 1976; Twarog 
1980a; Meusinger \etal 1991; Carraro \& Chiosi 1994). 
In the following, we 
will assume that these abundance variations are randomly distributed within 
the metallicity range observed at a given stellar age. 
This is particularly important when considering possible 
explanations for the observed stellar abundance variations in detail (see 
Sect. 4; \cf Wielen \etal 1996).

\

\begin{figure*}[htp]
\leftline{\psfig{figure=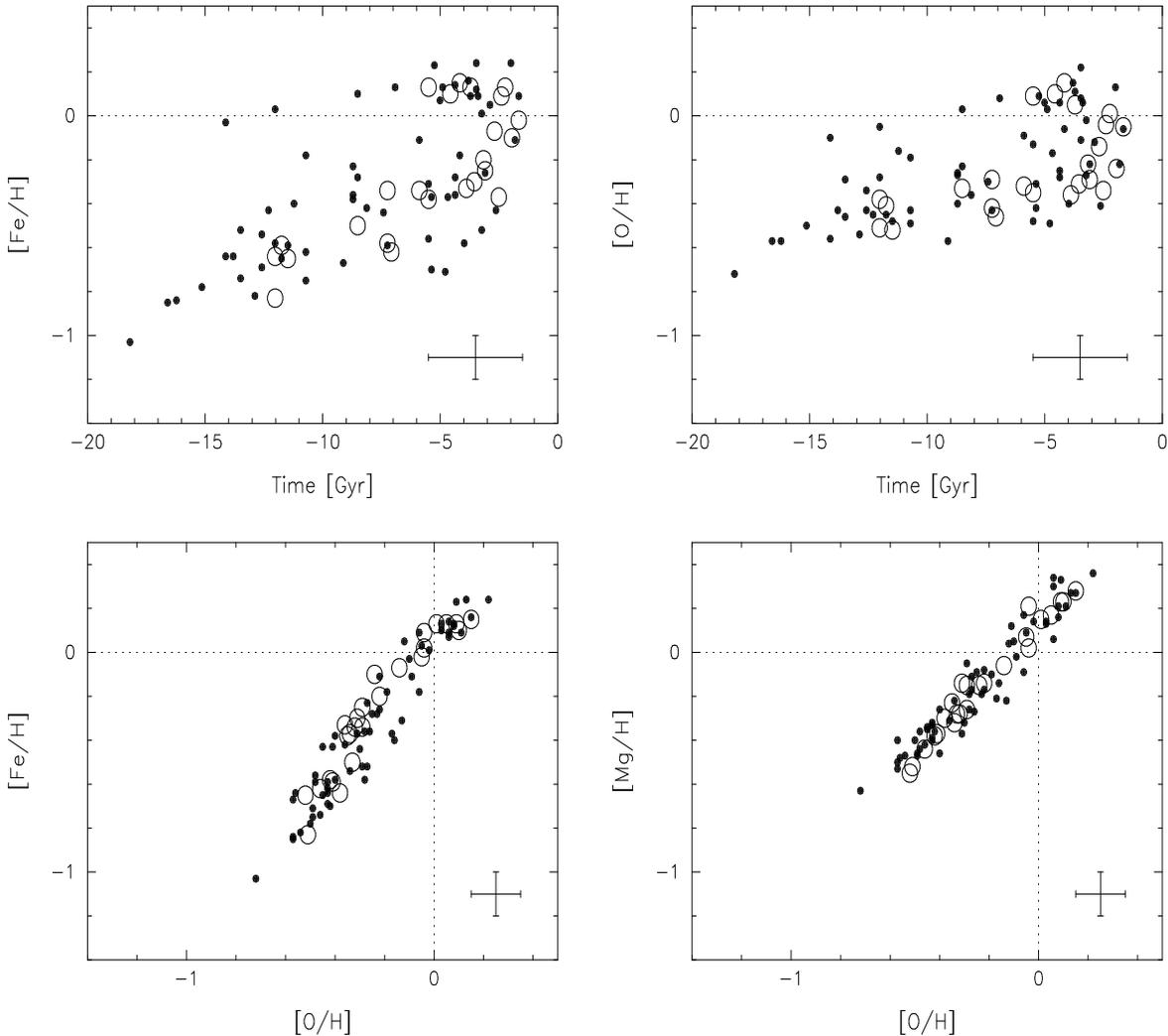,height=16.cm,width=18.cm,angle=270.}}
\vspace{-0.7cm} 
\caption[]{ Observed iron, oxygen, and magnesium abundance ratios
for main-sequence F and G dwarfs in the solar neighbourhood (data from
Edvardsson \etal 1993a).
Open circles represent stars with mean stellar galactocentric
distances at birth within 0.5 kpc from the Sun ($R_{\odot}$ = 8.4 kpc).
Full dots indicate stars with average distances within $\sim$2 kpc from the Sun.
Typical errors are indicated at the bottom right of each panel.
Note that the abundances of the most metal-poor disk stars included in this
sample resemble those of metal-rich halo dwarfs and giants (\eg Bessell \etal
1991; Gratton \& Sneden 1991; Nissen \etal 1994).
We assumed solar abundance ratios by number of $^{10}\log {\rm (O/H)}_{\odot} =
-3.13$, $^{10}\log {\rm (Fe/H)}_{\odot} = -4.51$,  and
$^{10}\log {\rm (Mg/H)}_{\odot} = -4.42$ and a hydrogen mass fraction in the
Sun of 0.68 (see Anders \& Grevesse 1989; Grevesse \& Noels 1993).
{\em Top panels}: Distributions of [Fe/H] (left) and [O/H] abundance ratios
vs. galactic age. {\em Bottom panels}: [Fe/H] vs. [O/H] (left) and
[Mg/H] vs. [O/H]}
\end{figure*}

\

Ages, abundances, and kinematical properties of 
the dwarfs belonging to the EDV sample are consistent with earlier 
investigations (\eg Twarog 1980a; Carlberg \etal 1985; Meusinger \etal 1991). 
An extensive discussion of the possible sources of errors in the abundance 
and age analysis as well as several consistency checks can be found in 
Edvardsson \etal (1993a,b). Errors due to data reduction uncertainties are 
estimated to lead to errors of at most 0.05 to 0.1 dex in abundance ratios
[M/Fe] as well as in [Fe/H] (see EDV). These errors are not expected to 
vary in a systematic manner with the derived stellar abundances and 
corresponding corrections will probably not reduce the observed variations.
Edvardsson \etal estimated errors in relative ages of $\sim$25\% for stars with 
similar abundances (absolute errors may be considerably larger). Thus, ages of 
stars as old as the Sun are estimated to be accurate within $\sim$1$-$2 Gyr. 
We conclude that errors in the abundances and ages of the sample stars are 
unlikely to account for the observed abundance variations, at least for stars 
with intermediate ages of $\sim$5 Gyr.

Knowledge of the formation sites of the sample stars is important to decide 
whether or not orbital diffusion in combination with radial abundance 
gradients in the Galactic disk can provide an adequate explanation for the 
observed abundance variations. 
Galactocentric distances of the sample stars {\em at birth} were 
obtained using stellar orbits reconstructed from their {\em present-day} 
galactocentric distances, proper motions, and radial velocities, and using 
both theoretical and empirical models for the Galactic potential as discussed 
by EDV. Accordingly, nearly 85\% of the sample stars were found to have mean 
galactocentric distances $R_{\rm m}$ at birth within 1 kpc from the Sun 
(assuming $R_{\odot} \sim 8.4$ kpc at present). However, predictions of the 
diffusion of stellar orbits in space,
based on the observed relation between velocity dispersion and age for nearby 
stars, suggest that many stars may have been formed at galactocentric 
distances as large as $\sim$4 kpc from where they are nowadays observed in the 
SNBH (Wielen \etal 1996). 
In either case, these nearby stars 
trace the evolution of the Galactic disk ISM over a much wider range in 
galactocentric distance then they are observed.

As an independent test to examine whether stellar orbital diffusion can be the 
main cause for the observed abundance variations, we translated stellar 
abundance deviations $\Delta$[M/H] from the mean abundances of similar aged 
stars born at $\sim R_{\odot}$ into galactocentric distance differences 
$R_{\rm m} - R_{\odot}$. This was done independently for [Fe/H] and [O/H] 
abundance ratios assuming present-day local radial abundance gradients 
of $-$0.07$\pm$0.015 dex kpc$^{-1}$ in [O/H] (\eg Shaver \etal 1983; 
Grenon 1987; see also Wilson \& Matteucci 1992) and $-$0.1 dex kpc$^{-1}$ in 
[Fe/H] (see \eg EDV). Clearly, distances $R_{\rm m}$ based on oxygen and iron 
abundances are expected to be similar (\eg $\Delta R_{\rm m}^{\rm O, Fe} \la$ 
1 kpc) when orbital diffusion is important for the observed stellar abundance 
variations.

In this manner, we find that $\Delta R_{\rm m}^{\rm O, Fe} \ga 1$ kpc for 47 
stars in the EDV sample (\ie $\sim$56\%). Similarly, we find $\Delta 
R_{\rm m}^{\rm O, Fe} \ga$ 2, 3, 
and 4 kpc, for $\sim$31, 14, and 8\% of the sample stars, respectively.
We note that the derived values of $\Delta R_{\rm m}^{\rm O, Fe}$ are 
insensitive to 
the stellar age but depend on the assumed radial abundance gradients as well as
on the mean [M/H] vs. age relations adopted for stars born at $R_{\odot}$.
Although a detailed investigation of the uncertainties involved is beyond the 
scope of this paper (\eg the variation of abundance gradients with disk age; 
\cf Grenon 1987), we estimate that $\Delta R_{\rm m}^{\rm O, Fe} \la$ 
0.8 (1.5) kpc for {\em typical} errors of 0.05 (0.1) dex in both [Fe/H] 
and [O/H] for most of the sample star (assuming a gaussian error distribution).
This suggests that a substantial part of the observed abundance variations 
is difficult to explain by stellar orbital diffusion only.

Also, Edvardsson \etal argued that the magnitude of the observed variations 
will reduce to $\Delta$[M/H] $\sim 0.3$ dex if one accounts properly for 
systematic errors and possible effects of stellar orbital diffusion. 
However, a reduction of the abundance spread among field dwarfs to 
$\Delta$[Fe/H] $\sim 0.3$ dex seems contradicted by the observed 
variations of $\Delta$[Fe/H] $\sim 0.5$ $\pm 0.1$ dex among similarly aged 
open clusters 
{\em after} correcting for radial abundance gradients across the Galactic plane 
(Carraro \& Chiosi 1994). Apart from this, such a reduction appears 
inconsistent with the observed abundance spread of [O/H] $\ga$0.4 dex among 
B stars {\em at a 
given galactocentric radius} between 7 and 16 kpc in the disk (Gehren \etal 
1985; Kaufer \etal 1994) and with the large abundance variations of $\sim$0.7 
dex observed among young open clusters over a distance scale of only
$\sim$1 kpc at a galactocentric radius of $\sim$13 kpc (Rolleston \etal 1994).

What fraction of the current disk stellar 
population actually formed in the Galactic halo depends on the 
detailed dynamical evolution of the disk which is not well known (\eg Pagel 
\& Tautvaisiene 1995). However, 
most stars in the EDV sample have derived maximum distances from the 
Galactic plane at birth of $h_{\rm max}$$<$0.5 kpc. This largely excludes 
halo stars from the sample and further implies that abundance gradients 
perpendicular to the Galactic plane are inadequate as explanation 
for the observed abundance variations (\eg Carney \etal 1990). 

From these arguments, we conclude that orbital diffusion of stars from 
elsewhere in the Galactic disk is probably insufficient as explanation for 
the observed variations in [Fe/H] and [O/H] among F and G dwarfs in 
the SNBH. This conclusion is consistent with the finding that abundance 
variations for subsamples of stars restricted to be born within 1 and 0.5 kpc 
from the Sun, respectively, are similar to those for the complete sample 
(see EDV; \cf Fig. 1). 
Therefore, we believe that differential chemical evolution and mixing of 
interstellar gas must be an important cause for the large stellar abundance 
variations observed in the SNBH as well. The abundances of the 
Sun and of open clusters in the Galactic disk fit well into this picture, 
as is argued below.

\subsection{Chemical evolution of the solar neighbourhood}

Detailed comparison of abundances within local H{\sc ii} regions and 
the Sun have shown that oxygen (among other heavy elements) is underabundant in 
the H{\sc ii} regions by about 0.15$-$0.3 dex (\eg Shaver \etal 1983; Peimbert 
1987; Baldwin et al. 1991; Osterbrock, Tran \& Veilleux 1992). 
Also, CNO-abundances of H{\sc ii}-regions and B main-sequence stars in the 
Orion nebula were found smaller than corresponding 
abundances in the Sun (Cunha \& Lambert 1992; Gies \& Lambert 1992). 
The remarkable result that the Sun is metal-rich by $\sim$0.15$-$0.2 dex in 
[O/H] compared to its surroundings is also supported by observations of B stars 
in nearby associations and young clusters (Fitzsimmons \etal 1990), diffuse 
interstellar clouds (\eg York \etal 1983), and disk planetary nebulae 
(de Freitas Pacheco 1993; Peimbert \etal 1993). Although abundance 
determinations in the SNBH may be biased towards regions associated with infall 
of metal-poor gas or suffer from heavy element depletion by dust, 
the existence of many metal-poor regions in the SNBH would be difficult to 
reconcile with efficient mixing in the local disk ISM (\eg Roy \& Kunth 1995).

The above observations are consistent with the Edvardsson \etal data which 
suggest that the Sun is metal-rich by 0.2$-$0.25 dex in [O/H] and by 
0.25$-$0.3 dex in [Fe/H] compared to the mean abundances of stars which formed 
in the SNBH $\sim$4.5 Gyr ago (\cf Fig. 1). These observations support
the idea that the Sun is metal-rich for its age (see also Steigman 1993) and 
that abundance inhomogeneities in the local disk ISM did exist. 
The fact that the Sun is metal-rich by a factor of $\sim$1.5$-$2 compared to 
nearby regions currently experiencing star formation may be explained by
self-enrichment of the gas cloud out of which the Sun was born
(\eg Gies \& Lambert 1992; Peimbert \etal 1993). 
Alternatively, orbital diffusion of the Sun may play an important 
role (Wielen \etal 1996).
We will discuss arguments in support of the former possibility in Sect. 5.2.

\subsection{Open clusters}

Variations in [Fe/H] among disk open clusters of a given age are 
known to be larger than any possible trend of [Fe/H] with age (\eg Nissen 1988; 
Boesgaard 1989; Garcia-Lopez \etal 1993; Friel and Janes 1993).
Recently, abundance variations of $\sim$0.5 $\pm 0.1$ dex in [Fe/H] among 
clusters of a 
given age {\em after} correcting for the radial abundance gradient across the 
Galactic plane have been reported by Carraro \& Chiosi (1994).
The observed abundance variations among open clusters appear
somewhat smaller (\ie by $\sim$0.2-0.3 dex in [Fe/H]) than those among 
field F and G dwarfs in the EDV sample.
However, the magnitude of the observed variations 
suggests that the processes responsible for the abundance inhomogeneities 
among field stars in the SNBH and among open clusters widespread 
throughout the Galactic disk may well be the same. 

The lack of a tight age-metallicity relationship for open clusters in the 
Galactic disk suggests that the chemical enrichment of the disk ISM has been 
inhomogeneous on time scales less than $\sim$10$^{8}$$-$10$^{9}$ yr, 
consistent with the abundance variations observed for intermediate age F and 
G dwarfs discussed above.

\section{Model characteristics and assumptions}

In the previous section, we have argued 
that differential chemical evolution and mixing of interstellar 
gas probably provides the main explanation for the large abundance variations 
observed among similarly aged stars in the SNBH. In this case, abundance 
inhomogeneities in the global disk ISM may result from local mixing of 
metal-deficient material (\eg infall) and/or local mixing of metal-enhanced 
material (\eg stellar enrichment). When star formation is initiated within 
the mixed material before any abundance fluctuations are wiped out, these 
inhomogeneities can be recorded by long-living stars. 

Efficient mixing by stellar winds and supernova 
explosions is generally accepted to occur within $\sim$10$^{7}-10^{8}$ yr
(\eg Edmunds 1975; Ciotti \etal 1991; Roy \& Kunth 1995). This suggests 
that the processes responsible for the onset of star formation and 
those causing substantial abundance inhomogeneities in the disk ISM 
are the same.
We consider this as a strong argument in favour of sequential star 
formation and/or infall induced star formation as the main processes 
responsible for the observed abundance variations (ample observational 
support for the occurence of these processes in the local disk ISM
are briefly discussed in Sect. 5.2). Obviously, the quantitative effect of 
these processes on stellar abundance variations, {\em relative} to the mean 
abundances in the local ISM, depends on the detailed chemical evolution 
of the disk ISM.

\subsection{Model description}

We present a model for the inhomogeneous chemical evolution of a star 
forming gas cloud. The basis for this model forms
the individual star formation history and chemical evolution 
of multiple subclouds that mutually exchange interstellar material.
We here restict ourselves to a brief outline of the basic assumptions and 
model characteristics. A more detailed description of the equations and 
input physics used is given in the (Appendix to the) electronic version of 
this paper.

We start from a homogeneous gas cloud with total mass M$_{\rm cl}$.
At any time the cloud is subdivided into ${\rm N}_{\rm scl}$ star forming,
active subclouds (with corresponding masses M$_{\rm scl}^{i}$) 
and a quiescent, inactive cloud part (with mass M$_{\rm qcl}$) not 
experiencing star formation. Each subcloud is allowed to follow its 
individual star formation, infall, and mixing history. 
Infall of matter is considered by allowing episodic mixing of metal-deficient 
material to each subcloud separately.

\

\begin{figure*}[htp]
\leftline{\psfig{figure=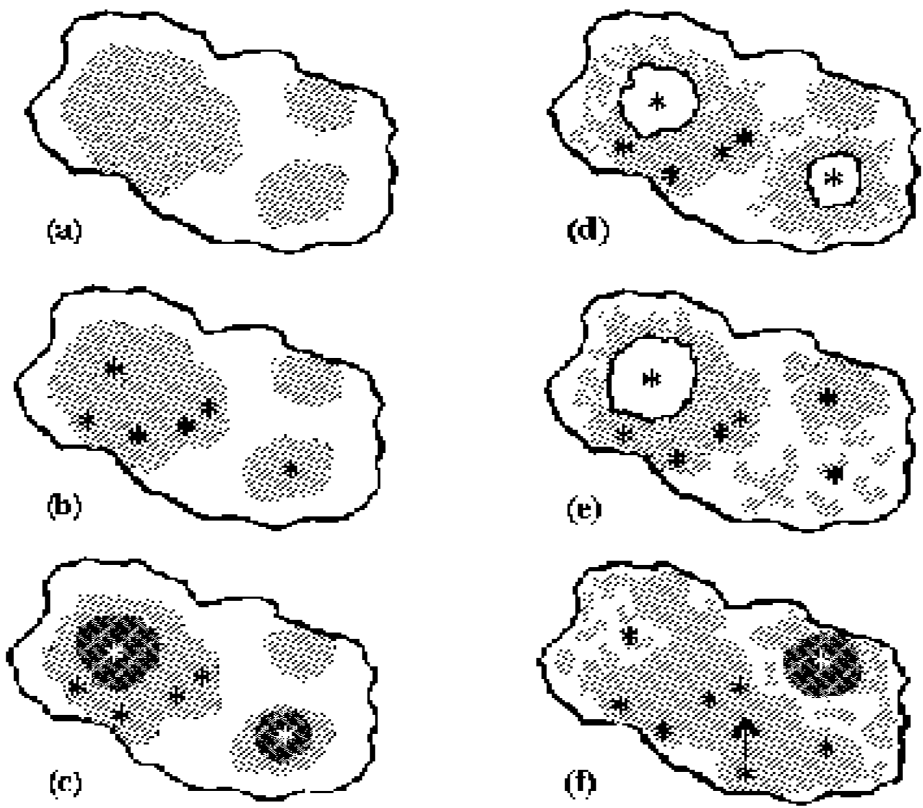,height=12.cm,width=13.cm,angle=270.}}
\vspace{-0.7cm} 
\caption[]{Schematic model for the inhomogeneous chemical evolution of a star
forming cloud: evolutionary sequence of star formation, enrichment and
mixing processes. Shown is a star forming cloud region.
Each of the processes indicated in this region may
occur in other regions of the cloud as well.
Symbols have the following meaning:
{\bf a} subclouds indicated by hatched areas, {\bf b} star formation indicated
by asterisks, {\bf c} stellar enrichment shown as shaded areas enclosing
white asterisks, {\bf d} subcloud core dispersal indicated as blanked out area
surrounding stars, {\bf e} break up of entire subcloud and initiation of star
formation in a nearby subcloud, {\bf f} arrow indicates stars entering a
subcloud from elsewhere. Each of the processes
indicated may occur frequently during the cloud evolution time $t_{\rm ev}$}
\end{figure*}

The adopted set of processes that modify the 
distribution of gas and stars within a star forming region of a molecular cloud
are illustrated in Fig. 2. Different 
subfigures refer to the following processes:
\newcounter{bobcnt}
\begin{list}{(\alph{bobcnt})}%
{\usecounter{bobcnt} \itemsep 0.cm \parsep 0.cm \leftmargin 0.4cm 
\itemindent 0.55cm}
\item subcloud formation from the inactive cloud ISM (and/or 
from infalling material);
\item conversion of gas into stars (star formation at distinct subcloud cores); 
\item ejection of material by stars to their immediate surroundings;
\item mixing of dispersed core material with subcloud after star formation 
event;
\item break up of entire subcloud, mixing with inactive cloud ISM, and 
induced star formation;
\item enrichment of subcloud by stars not formed within the subcloud.
\end{list}
In our model, the inhomogeneous chemical evolution of a star forming gas cloud, 
consisting of many subclouds, is determined by the combined effect of the 
above processes. 
During a time-interval $\Delta t$, these processes may occur 
simultaneously within each subcloud.
In this manner, the initial abundances of a newly formed stellar generation 
are determined by: 1) the enrichment of the subcloud 
by preceding stellar generations, and 2) the mixing 
history of the subcloud with the ambient ISM. 

In brief, the adopted evolution scenario is as follows.
Subcloud formation (Fig. 2a) is assumed to occur either from the inactive 
cloud ISM and/or from infalling material (details related to the infall model 
will be given in Sect. 4.3).
During the lifetime $t_{\rm ev}$ of the entire system, a total number 
of $N_{\rm sf}$ star formation events is assumed to occur. Each star 
formation event is assumed to take place in an active subcloud (Fig. 2b).
Each subcloud is allowed to experience numerous star formation events 
and/or to remain inactive during a substantial part of its lifetime.
Consequently, each subcloud can be enriched by one or multiple star formation 
events dictating its chemical evolution (Fig. 2c). 
When the active subcloud {\em core} is dispersed by stellar winds and/or 
supernova shocks, part of the enriched matter is assumed to mix 
homogeneously with the surrounding subcloud material (Fig. 2d). 
The remaining part is 
assumed to mix homogeneously either to a nearby subcloud hosting the {\em next} 
star formation event or to the ambient inactive cloud part (Fig. 2e).
No mass-exchange is assumed between the subcloud and the ambient inactive 
cloud ISM during the time interval in which two or more star formation events 
occur within the same subcloud.
In addition, subcloud material may be enriched by stars formed outside
the subcloud. In this case, stars from elsewhere in the inactive cloud
occasionally enter the subcloud region and enrich the subcloud by means of 
their ejecta (Fig. 2f).
Stellar enrichment by old stellar generations is assumed to proceed 
continuously with time but is considered in detail only at specific 
evolution times corresponding to the occurence of any of the
discontinuous processes referred to in Fig. 2.

We define the subcloud core dispersal time $\Delta t_{\rm disp}$ as the time 
between onset of star formation within a subcloud core region and the 
complete dispersal of this region.
This time interval constraints the mass of the most massive 
star that is able to enrich the subcloud core material before the core 
ultimately breaks up.
Before dispersal of an {\em entire} subcloud, the subcloud will be
enriched by the stellar populations it is hosting. After subcloud 
dispersal (\ie after a typical mixing time scale $\Delta t_{\rm mix}$), 
stars and gas belonging to the subcloud are assumed to mix instantaneously 
and homogeneously with the inactive cloud ISM. 
Subsequently, different cloud fragments may combine to form  
new subclouds wherein star formation occurs as soon as the critical conditions 
for star formation are met.
The mixing history of each subcloud determines the inhomogeneous chemical 
evolution of the inactive cloud part as well as that of nearby subclouds.
For simplicity, we do not consider partial mixing of subcloud 
material to the inactive cloud. 

\subsection{Outline of model computations}

We perform Monte-Carlo simulations of the 
inhomogeneous chemical evolution of a star forming gas cloud.
The continuous process of formation and break up of subclouds and 
of the formation and dispersion of subcloud core regions associated 
with star formation, are followed as outlined in the previous 
section. 
During the evolution of the cloud, we keep track of the total 
mass contained in gas and stars as well as the stellar and interstellar 
abundances of H, He, C, O, Fe, Mg, Al, and Si, both within 
each subcloud and the inactive cloud part. 
No instaneous recycling is assumed, \ie metallicity dependent stellar 
lifetimes are taken into account. 

\subsection{Model input parameters}

Model input parameters for the reference model are listed in Table 1.
We distinguish parameters related to: 1) the entire cloud and inactive cloud 
part, 2) active subcloud regions, and 3) individual star formation events: 

\noindent $\bullet$ {\em Cloud and inactive cloud part:} 
The initial cloud mass $M_{\rm cl}$ is treated as a mass scaling parameter 
(\ie resulting abundances are not altered for different values of 
$M_{\rm cl}$).
We here adopted $M_{\rm cl}$ = 5 10$^{10}$ \ms similar to that of the Galactic 
disk (\eg Binney \& Tremaine 1987).
We assume a cloud evolution time $t_{\rm ev}$ = 14 Gyr.
This is comparable to the age of the Galaxy as derived from the 
age of the oldest globular clusters, \ie 14$\pm$3 Gyr (\eg Buonanno \etal 1989).
In our model, the impact of processes causing stellar abundance 
variations does not depend on the specific age of the Galactic disk assumed.

We consider a total number of star formation events during the cloud evolution 
time of typically $N_{\rm sf}=100$. In practice, $N_{\rm sf}$ is limited only 
by the preferred model run time, \ie 1-2 hours on a HP Apollo 715 machine.
The total number of {\em subclouds} $N_{\rm scl}$ is determined by the 
number of star formation events within each subcloud.
For the reference model, we assume a maximum number of star formation 
events within one subcloud $N_{\rm sf}^{\rm max}$ = 1 
so that $N_{\rm scl}$ = $N_{\rm sf}$.
Cloud initial abundances $X_{\rm qcl}$ are as given in Table 1.
Initially, the cloud is considered homogeneous, metal-free, and void of stars. 

\noindent $\bullet$ {\em Active subclouds:} 
In case of the reference model,
we force subclouds to form at regular intervals of $t_{\rm ev}$ / 
$N_{\rm scl}$ = 1.4 10$^{8}$ yr. 
We assume the subcloud mass $M_{\rm scl}$ directly proportional to 
the entire cloud gas-to-total mass-ratio $\mu$ at time of subcloud formation 
$t_{\rm scl}$. This implies more massive subclouds to form at relatively 
high gas fractions $\mu(t)$. Assuming a constant star formation 
efficiency, this results in an exponential decrease of subcloud mass with 
disk age $t$ (\eg Clayton 1985):
\begin{equation}
M_{\rm scl} = M_{\rm scl}(0) \exp (-t / t_{\rm dec}) 
\end{equation}
where $M_{\rm scl}(0)$ is the mass of a subcloud formed at $t$=0 (which 
may vary between different models) and $t_{\rm dec}$ a characteristic time 
scale at which the mass of subsequent subclouds formed is assumed to decay
(identical for all models). The assumption of a star formation rate (SFR) 
directly proportional to the subcloud formation rate is not essential for the 
results discussed here. 

The decay time scale $t_{\rm dec}$ is constrained observationally by the ratio 
of the average past to present SFR in the Galactic disk ($\sim 3-$7; \eg 
Mayor \& Martinet 1977; Dopita 1990). 
We here assume an exponentially decaying SFR with $t_{\rm dec} =$ 6 Gyr.  
As will be shown in Sect. 4.1, this SFR can account simultaneously for the 
actual gas-to-total mass-ratio in the disk of $\mu_{1} \sim 0.05-$0.2 
(Kulkarni \& Heiles 1987; Binney \& Tremaine 1987; see also Basu \& Rana 1992), 
the smooth increase in the global AMR for elements such as O and Fe, and the 
magnitude of the current SFR in the Galactic disk (\ie $\sim$3.5 M$_{\odot}$ 
yr$^{-1}$; \eg Dopita 1987). In contrast, constant SFR models are inconsistent 
with these observations (Twarog 1980a; see also Clayton 1985).

The time between the formation and complete mixing of a subcloud to the 
inactive cloud part is defined as $\Delta t_{\rm mix}$. This time scale 
has been considered to allow for the individual chemical evolution of a 
subcloud isolated from the inactive ISM (see below).

\noindent $\bullet$ {\em Individual star formation events:}
We assume the onset of star formation within each subcloud to coincide with 
the formation of the subcloud itself in case of the reference model.
This results in a grid of regularly spaced star formation times 
$t_{\rm sf}$ = $t_{\rm scl}$.

We define the core dispersal time $\Delta t_{\rm disp}$ as the time between
onset of star formation $t_{\rm sf}$ within a subcloud core and the moment 
star formation ends due to the actual break up of this core.
Observational estimates of this time scale are generally $\la$10$^{7}$ yr 
(\eg Garmany \etal 1982; Leisawitz 1985; Genzel \& 
Stutzki 1989; Rizzo \& Bajaja 1994; Haikala 1995). 
For the reference model, we assume the 
{\em entire} subcloud to break up at time of dispersal of the star forming  
subcloud {\em core}, \ie $\Delta t_{\rm mix}$ = $\Delta t_{\rm disp}$. 

The star formation efficiency $\epsilon^{j}$ is defined as the amount of 
subcloud matter $\Delta M_{\rm scl}$ turned into stars during star formation 
event $j$. In fact, the star formation efficiency determines the 
amount of material to which the stellar ejecta of a previous stellar 
generation are mixed within a given star forming cloud.
Observational estimates for $\epsilon$ in molecular clouds 
in the Galactic disk span a wide range: between a few tenths of a percent to 
$\sim$50\% (\eg Wilking \& Lada 1983). We will discuss the values 
assumed for $\epsilon^{j}$ in Sect. 4.

Sequential stellar enrichment is taken into account by assuming that
a fraction $\lambda^{j}$ of enriched material associated with star 
formation event $j$ is mixed homogeneously to the subcloud hosting 
the next star formation event. 
Subcloud material not converted into stars is  
mixed to the inactive cloud part after complete dispersal of the 
subcloud. The relative importance of these model parameters on the 
resulting stellar abundance variations will be discussed in Sect 4.2.

\begin{table*}[htp]
\caption[]{List of input parameters (values listed for reference model)}
\begin{flushleft}
\begin{tabular}{lll}
\multicolumn{3}{l}{\em Cloud and inactive cloud part}  \\ \hline
$M_{\rm cl}$  & 5 10$^{10}$ \ms &
total cloud total mass \\
$t_{\rm ev}$  & 14 Gyr              & cloud evolution time \\
$N_{\rm sf}$  & 100                 & total number of star formation events \\
$N_{\rm scl}$ & 100                 & total number of subclouds formed \\
$X_{\rm qcl}(0)$ & H=0.76 & initial cloud abundances; He=0.24
\\ \hline
& & \\
\multicolumn{3}{l}{\em For each subcloud i} \\ \hline
$t_{\rm scl}$        & 1.4 10$^{8}$ yr & time at which subcloud is formed, \ie
each $t_{\rm ev}$ / $N_{\rm scl}$ = 1.4 10$^{8}$ yr \\
$M_{\rm scl}$        & exp & exponentially decaying subcloud mass at time of
formation ($M_{\rm scl}(t=0)$ = 6 10$^{9}$ \mss) \\
$N_{\rm sf}^{\rm max}$   & 1 & maximum number of SF events
within one subcloud \\
$\Delta t_{\rm mix}$ & $\Delta t_{\rm disp}$ & mixing time of entire subcloud
\\ \hline
& & \\
\multicolumn{3}{l}{\em For each star formation event j} \\ \hline
$t_{\rm sf}$ & $t_{\rm scl}$ & evolution time at which SF-event $j$ occurs \\
$\Delta t_{\rm disp}$ &  10$^{7}$ yr & subcloud core dispersal time \\
$\epsilon$  & 0.50 & subcloud star formation efficiency$^{(1)}$ \\
$\lambda$ & 0 & efficiency of sequential enrichment$^{(2)}$ \\ \hline
\end{tabular}
\end{flushleft}
\parbox{12.3cm}{\begin{small} {\bf Notes}: ({\bf 1}) $\epsilon^{j}$ is the
mass fraction of the subcloud converted into stars during dipsersal time
$\Delta t^{j}_{\rm disp}$,
({\bf 2}) $\lambda^{j}$ refers to the amount of stellar material
returned to the subcloud core hosting the {\em next} star formation event.
\end{small}}
\end{table*}

\begin{table*}[htp]
\caption[]{IMF related parameters and stellar enrichment}
\begin{flushleft}
\begin{tabular}{lll}
\hline
$\gamma$    & --2.35      & slope of power-law IMF $m^{\gamma}$ \\
$m_{\rm l}$, $m_{\rm u}$ & 0.1, 60 \ms  & stellar mass limits at birth \\
$m_{\rm l}^{\rm SNII}$, $m_{\rm u}^{\rm SNII}$ & 8, 30 \ms & progenitor
mass range for SNII and SNIb/c \\
$m_{\rm l}^{\rm SNIa}$, $m_{\rm u}^{\rm SNIa}$ & 2.5, 8 \ms & progenitor
mass range for SNIa \\
$\phi^{\rm SNIa}$       & 0.005 & fraction of progenitors ending as SNIa \\
$\phi^{\rm SNIb/c}$     & 0.33 & fraction of SNII progenitors ending as SNIb/c
\\ \hline
\end{tabular}
\end{flushleft}
\end{table*}

\subsection{Stellar evolution data}

We follow the stellar enrichment of the star forming cloud in terms of the 
characteristic element contributions of Asymptotic Giant Branch (AGB) stars, 
SNII, SNIa, and SNIb/c. This treatment is based on the specific abundance 
patterns observed within the ejecta of each of these stellar groups 
(\eg Trimble 1991; Groenewegen \& de Jong 1992; van den Hoek \etal 1996).
We take into account metallicity dependent stellar element yields, 
remnant masses, and ages, while assuming the stellar ejecta to be returned 
at the end of the lifetime of the star (see \eg Maeder 1992; Schaller \etal 
1992). The respective time delays in enrichment by SNIa and SNII are accounted 
for in detail. A more detailed description of the combined set 
is given elsewhere (\eg van den Hoek \etal 1996; see also electronic 
version of this paper). 

For AGB stars (initial mass $m\la 8$ \mss) we adopt the metallicity 
dependent yields presented by Groenewegen \& de Jong (1992). These yields are 
based on a synthetic evolution model for AGB stars and are succesful in 
explaining the observed abundances in carbon stars and planetary nebulae in 
the Galactic disk (Groenewegen \etal 1995). 
For Type-II SNe, we use the explosive nucleo-synthesis yields (independent of 
initial metallicity) described in detail by Hashimoto \etal 
(1993) and Thielemann \etal (1993) for stars with 8$\la$ $m$[\mss] $\la$60. 
The 20 \ms model of this set accounts well 
for the observed abundances in SN1987A (Nomoto \etal 1991). 
Explosive nucleo-synthesis yields for Type-Ia SNe are adopted from 
Nomoto \etal (1984; model W7 for SNIa at $Z={\rm Z}_{\odot}$ and $Z=0.0$ of 
the accreted material; see also Yamaoka 1993) and for SNIb/c from Woosley 
\etal (1995). According to these yields, typical amounts of iron produced are 
$\sim 0.08$ \ms for SNII, $\sim$0.8 \ms for SNIa, and $\sim$0.1 \ms for SNIb/c.

The adopted yields for SNIa, b/c are relatively uncertain 
due to unknown details of the progenitor history and
the explosion mechanism (either binary or single 
star evolution; see \eg Smecker-Hane \& Wyse 1992; Woosley \etal 1993).
However, we do not believe that these uncertainties are relevant for the 
qualitative results obtained in this paper (\cf Sect. 5.2).

Metallicity dependent stellar yields for stars during their wind 
(\ie pre-SN) phase have been 
adopted from Maeder (1992, 1993), to whom we refer the reader also 
for a definition of the stellar element yields as used in the Appendix.
For stars with $m \ga 20$ \ms we used the higher mass loss rates in case 
$Z$=0.02 (\cf Maeder 1992; Schaller \etal 1992).
The mass $m_{\alpha}$ of the helium core left at the end of the He-burning 
phase (or C-burning phase for massive stars) has been used as input for the 
SNII and SNIb/c nucleosynthesis models referred to above. Yields were 
linearly interpolated both in $m$ and $m_{\alpha}$. Errors due to the 
coupling of these sets of stellar evolution data are probably small and 
are neglected here (\cf van den Hoek \etal 1996).
Remnant masses and stellar lifetimes were adopted from 
the Geneva group as well (\eg Schaller \etal 1992).

For the reference model, the adopted IMF-slope, stellar mass limits at birth, 
and progenitor mass ranges for stars ending their lives as SNIa and 
SNII($+$SNIb/c) are listed in Table 2.
Stars with $m > 60$ \ms have been excluded because their theoretical yields are 
rather uncertain (\eg Maeder 1992). We expect that the IMF-weighted 
contribution by such stars to the enrichment of the ISM is relatively low.
Stars more massive than $m_{\rm u}^{\rm SNII}$ are assumed not to
explode as supernova but to end as black hole 
(\cf Maeder 1992; Nomoto \etal 1994; Prantzos 1994; Tsujimoto \etal 1995). 
Consequently, stars with $m \ga m_{\rm u}^{\rm SNII}$ contribute to the ISM 
enrichment during their stellar wind phase only. When no upper mass limit 
$m_{\rm u}^{\rm SNII}$ = 25$-$30 \ms is introduced, models using 
up-to-date SNII yields predict abundances that are too high compared to those 
observed in the ISM, in particular for helium and oxygen (\eg Twarog \& 
Wheeler 1982; Maeder 1992, 1993; Timmes \etal 1995). 

We consider a fraction $\phi^{\rm SNIa} = 0.005$ of all WD
progenitors with initial mass between $\sim 2.5$ and 8 \ms to end as SNIa. 
A more detailed discussion of the contribution by SNIa is postponed to Sect. 
5.1 (see also Ishimaru \& Arimoto 1995).
In addition, we assume about one third of all supernova progenitors with 
$m_{\rm l}^{\rm SNII}$ $\la$$m$$\la$ $m_{\rm u}^{\rm SNII}$ 
to end as SNIb/c, \ie $\phi^{\rm SNIb/c} \equiv$ [SNIb/c / (SNIb/c + SNII)] = 
0.33. This value is based on the observed ratio of current formation rates of 
SNII and SNIb/c in the Galaxy (\eg van den Bergh \& Tammann 1991; Tutukov, 
Yungelson \& Iben 1992; Cappellaro \etal 1993).

\begin{figure*}[ht]
\leftline{\psfig{figure=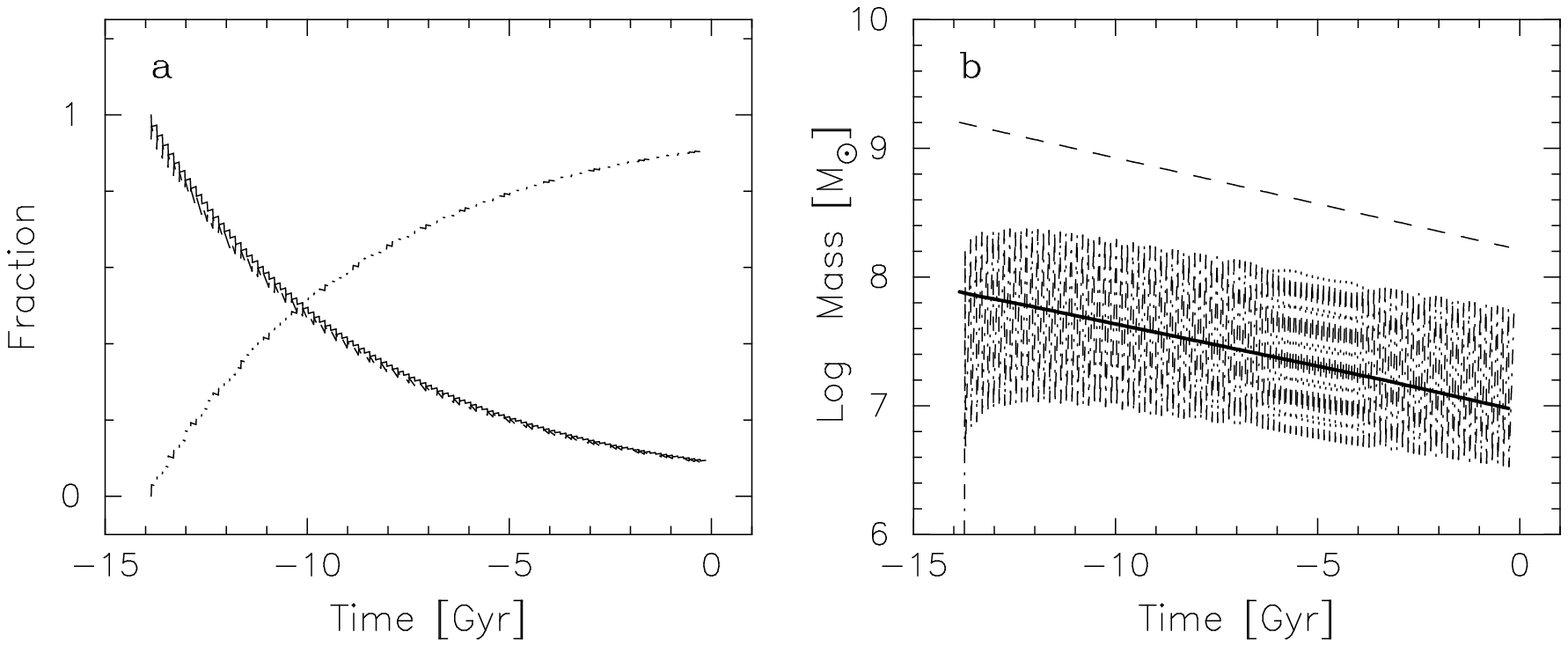,height=5.cm,width=14.cm,angle=270.}}
\vspace{0.5cm} 
\caption[]{Reference model:
{\bf a} Stellar-to-total (dashed curve) and gas-to-total
(solid line) mass-ratios vs. age.
The gas-to-total mass-ratio for the inactive cloud part coincides
with that for the entire cloud,
{\bf b} Subcloud mass $\Delta M_{\rm scl}$ = $\epsilon M_{\rm scl}$
converted into stars (dashed curve) and amount of
gas returned to the subcloud by newly formed stars within $\Delta t_{\rm sf}$
(solid curve) vs. age. We assumed $M_{\rm scl}(t=0)$ = 6 10$^{9}$ \mss.
Total mass returned by previously formed
stellar populations present in the inactive cloud is shown for
comparison (dash-dotted curve). Fluctuations in this curve result from
integration over different time intervals, \ie $\Delta t_{\rm sf}$ and
$t_{\rm scl}$ (\cf Table 1).}
\end{figure*}

\section{Results}

We present results for the inhomogeneous chemical evolution model described in 
the previous section.
First, we consider the reference model which does not incorporate stellar 
abundance variations at a given age.
Thereafter, we discuss models that do incorporate stellar abundance variations 
due to: 1) sequential stellar enrichment, 2) infall of metal-deficient matter, 
and 3) combined infall of metal-deficient matter and sequential enrichment. 

\subsection{Reference model}

We consider a homogeneous gas cloud with initial conditions as listed in 
Tables 2 and 3. Within this cloud, active subclouds are formed at regular 
time intervals of 1.4 10$^{8}$ yr so that in total N$_{\rm cl}$ = 100 
subclouds form during cloud evolution time $t_{\rm ev}$ = 14 Gyr. 
We assume no time-delay between the formation of the subcloud and the actual 
onset of star formation within that subcloud, \ie $t_{\rm sf} = t_{\rm scl}$, 
and further assume each subcloud to experience a single star formation event. 
During this event, lasting $\Delta t_{\rm disp}$ = 10$^{7}$ yr, half of the 
subcloud mass is converted into stars, \ie $\epsilon = 0.50$. After each event, 
both gas and stars contained within the subcloud are mixed homogeneously to 
the inactive cloud part. We note that the 
assumption of $\epsilon = 0.50$ has no physical meaning here other than 
defining the gas consumption rate as a function of cloud age.
Model related quantitites are given in Table 3 (see Sect. 5.2).

Figure 3a shows resulting stellar and gas-to-total mass-ratios vs. age 
for the reference model. According to the assumed variation of subcloud mass 
with cloud evolution time (\cf Sect. 3.2), the gas-to-total mass-ratio 
decreases exponentially from $\mu_{\rm cl}=$ 1 to 0.1 (corresponding 
decrease in subcloud mass converted into stars is shown in Fig. 3b).
Figure 3b illustrates that the amount of gas returned by massive 
stars during each star formation event is less than $\sim$5\% of the total 
amount of gas converted into stars during the same event. This ratio 
is determined primarily by $t_{\rm sf}$ and $\epsilon$ (see below). 
The amount of gas returned during $\Delta t_{\rm disp}$ by the {\em entire} 
stellar population within the inactive cloud part vs. cloud age  
is plotted for comparison.

\begin{figure*}[ht]
\leftline{\psfig{figure=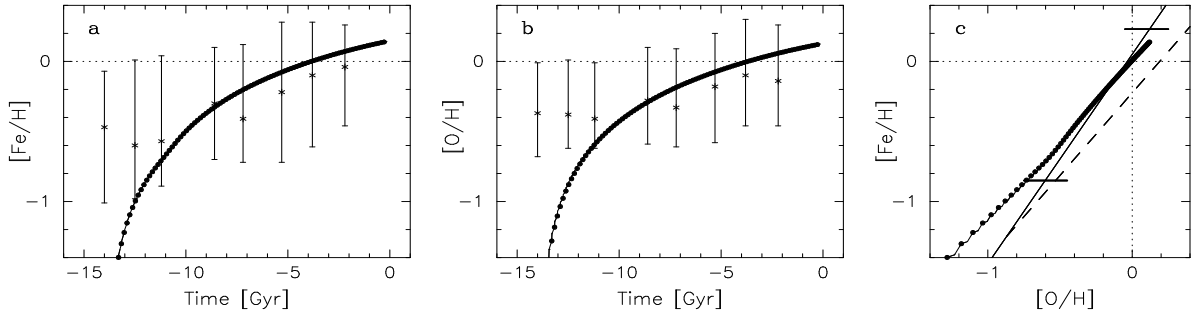,height=5.cm,width=16.cm,angle=270.}}
\vspace{-0.2cm} 
\caption[]{Reference model: Stellar [Fe/H] and [O/H]
abundance ratios vs. age. Stellar and ISM abundances at a given age are
exactly the same.
{\bf a} [Fe/H]: stellar abundances at birth (full circles) coinciding
with ISM abundances (solid curve).
Each full circle represents a stellar generation with
total mass of approximately $\Delta M_{\rm scl}$.
Asterisks with error bars indicate the maximum stellar abundance variations
observed in the Edvardsson \etal (1993a) data for main-sequence F and G
dwarfs (averaged over 1.5 Gyr bins). Data for stars older than 15 Gyr have
been omitted because of incompleteness (see Sect. 2).
{\bf b} [O/H]: curves and data similar to those for [Fe/H]. {\bf c}
Theoretical [Fe/H]
vs. [O/H]-relations including SNIa (full circles) and excluding SNIa (dashed
curve). Mean [Fe/H] vs. [O/H] relation for the Edvardsson \etal data
is shown as a straight line (horizontal marks indicate the observational range)}
\end{figure*}

Corresponding stellar and interstellar [Fe/H] and [O/H] abundance ratios 
are shown in Figs. 4a and b, respectively.  At a given age, stellar and ISM 
abundances are exactly the same so that abundance inhomogeneities do not 
occur. Note that the resulting AMRs do not depend on 
the adopted value for M$_{\rm cl}$ as long as the normalisation 
of the SFR remains such that the condition of a current gas-to-total 
mass-ratio $\mu_{1}$ of 0.1 is met.
The reference model predicts [Fe/H] and [O/H] abundance ratios that are 
consistent 
with the mean EDV data for stars younger than $\sim$ 10 Gyr. For stars 
older than 10 Gyr, agreement with the observations may be improved \eg 
by considering cloud ages in excess of $t_{\rm ev}$ = 14 Gyr or by detailed 
modeling of the halo-disk enrichment at early epochs of Galaxy evolution. 
We here concentrate on the stellar abundances observed during 
the last 10 Gyr of Galactic disk evolution.

\begin{table*}[htp]
\caption[]{Summary of model input parameters and resulting quantities
related to the SFR ($^{*}$scaled to $M_{\rm cl}$ = 2 10$^{11}$ \mss)}
\begin{flushleft}
\begin{tabular}{llllllrllll}
\hline
Model & Fig. & $M_{\rm scl}(0)$ & $N_{\rm sf}$ & $\epsilon^{\rm max}$ &
$m_{\rm u}^{\rm SNII}$ & $\mu_{1}$ & SFR$_{1}$ & INF$_{1}$ & $R^{\rm SNII}$ &
$R^{\rm SNIa}$ \\
  & & [\ms] & & & [\ms] & & [\ms yr$^{-1}$] & [\ms yr$^{-1}$] &
  [yr$^{-1}$] & [yr$^{-1}$] \\ \hline
Reference   & 3+4 & 1.3 (10) & 100 & 0.50 & 30 & 0.09 & 2.9 & $-$ &
       1.3 (-2) & 7.1 (-4) \\
Sequential (single) & 5-1, 5-2 & 2.1 (10) & 100 & 0.95 & 25 & 0.11 & 3.7 & $-$ &
			1.5 (-2) & 8.2 (-4) \\
Sequential (multiple)  & 5-3    & 1.5 (10) & 200 & 0.50 & 25 & 0.21 & 2.5 & 
$-$ & 1.0 (-2) & 6.4 (-4) \\
Infall     & 6    & 6.3 (9)  & 100 & 0.95 & 40 & 0.16 & 5.4 & 3.1 &
					  2.5 (-2) & 1.6 (-3) \\
Infall+Sequential    & 7-1, 7-2 & 6.4 (9)  & 136 & 0.90 & 25 & 0.19 & 5.2 & 
1.8 &
2.1 (-2) & 1.3 (-3) \\
Infall+Sequential   & 7-3 & 6.4 (9)  & 166 & 0.90 & 25 & 0.32 & 4.4 & 3.1 &
							    1.8 (-2) & 
1.1 (-3) \\
Observations$^{*}$ & & & & & & 0.05$-$0.2 & 3.6$\pm$1. & $\ga$0.5 &
	    2$-$4 (-2) & $\ga$ 3 (-3) \\ \hline
\end{tabular}
\end{flushleft}
\parbox{15.5cm}{\begin{small} $^{*}${\bf References}: \\
M$_{\rm cl}$: Bahcall \& Soneira (1980); Fich \& Tremaine (1991). \\
$\mu_{1}$: Kulkarni \& Heiles (1987); Binney \& Tremaine (1987); see also
Basu \& Rana (1992). \\
SFR: Dopita (1987);
Walterbos (1988; based on IR observations); Mezger 1988 \\
INF: Mirabel \& Morras (1984); L\'{e}pine \& Duvert (1994). \\
SNII \& SNIa: van den Bergh \& Tammann (1991); Tutukov \etal (1992);
Cappellaro \etal (1993); Strom (1993).
\end{small}}
\end{table*}

Our adopted values of $\phi^{\rm SNIa} = 0.005$ and $m_{\rm u}^{\rm SNII}$ = 
30 \ms provide optimal consistency with the mean observed [Fe/H] vs. [O/H] 
relation (\cf Fig. 4c). Clearly, the slope of the resulting [Fe/H] vs. [O/H] 
relation in case of enrichment by SNII only ($m_{\rm u}^{\rm SNII}$ = 60 \mss) 
is inconsistent with the observations. Thus, the data provided by 
Edvardsson \etal imply that SNII and SNIa,b/c nucleo-synthesis sites mixed 
their products together well. In addition, dilution of the
supernova ejecta by more metal-deficient material is needed to comply with the 
range in [Fe/H] and [O/H] observed for F and G dwarfs in the SNBH. 
This is simply because theoretically predicted (lifetime-integrated) mean [Fe/H]
ratios within the ejecta of supernova progenitors are in general much larger 
than those observed for long-living stars in the SNBH (see Sect. 3.4). 
Thus, whatever process is reponsible for the observed stellar abundance 
variations, both mixing of ejecta from different SN-types and dilution 
with metal-deficient material are involved. 

The resulting well-defined tight AMRs for the reference model are 
similar to those predicted by conventional single-zone chemical evolution 
models (\eg Twarog 1980a; Tinsley 1980).
Such models account for the global chemical enrichment
of the Galactic disk ISM during the last 10 Gyr, at least for elements like O 
and Fe, but they obviously provide no explanation for the observed variations 
in stellar abundances at a given age. 

\subsection{Sequential stellar enrichment}

In case of sequential star formation, efficient self-enrichment of a star 
forming gas cloud by successive stellar generations may result in  
abundance enhancements relative to the abundances in the ambient ISM.
When the local mixing time scale is larger than the time between two 
successive star formation events in such a cloud, these abundance enhancements 
can be deposited and recorded by newly formed stars.

In our model, the impact of sequential enrichment on abundance 
inhomogeneities in the ISM is determined by: a) the dispersal time of the 
star forming 
region, b) the total number of stellar generations formed within one and the 
same cloud, c) the efficiency of sequential enrichment, 
\ie the mass-ratio of the enriched stellar material and the cloud to which 
this material is mixed, d) details of stellar enrichment: \eg the 
relative number of SNII and SNIa, and e) the IMF and stellar mass limits at 
birth. We distinguish the effect of single and multiple sequential stellar 
enrichment on the stellar abundance variations. We will refer to single 
sequential enrichment as the case in which a star formation event induces 
subsequent star formation in a nearby cloud (when mixing enriched 
material to this cloud).

\subsubsection{Single sequential stellar enrichment}

We present results for models incorporating single sequential stellar 
enrichment (N$_{\rm sf}^{\rm max} $= 1). For illustration purposes, we 
consider only the alternating half of the subclouds to experience sequential 
enrichment. 
We define initial masses of subclouds that are sequentially enriched as 
$M_{\rm scl} \equiv \vartheta M_{*}$ where $M_{*}$ is the total mass of gas 
converted into stars during the previous enriching star formation event.
Initial masses of subclouds not involved with sequential enrichment 
are assumed to decrease exponentially (as for the reference model). 

\begin{figure*}[htp]
\leftline{\psfig{figure=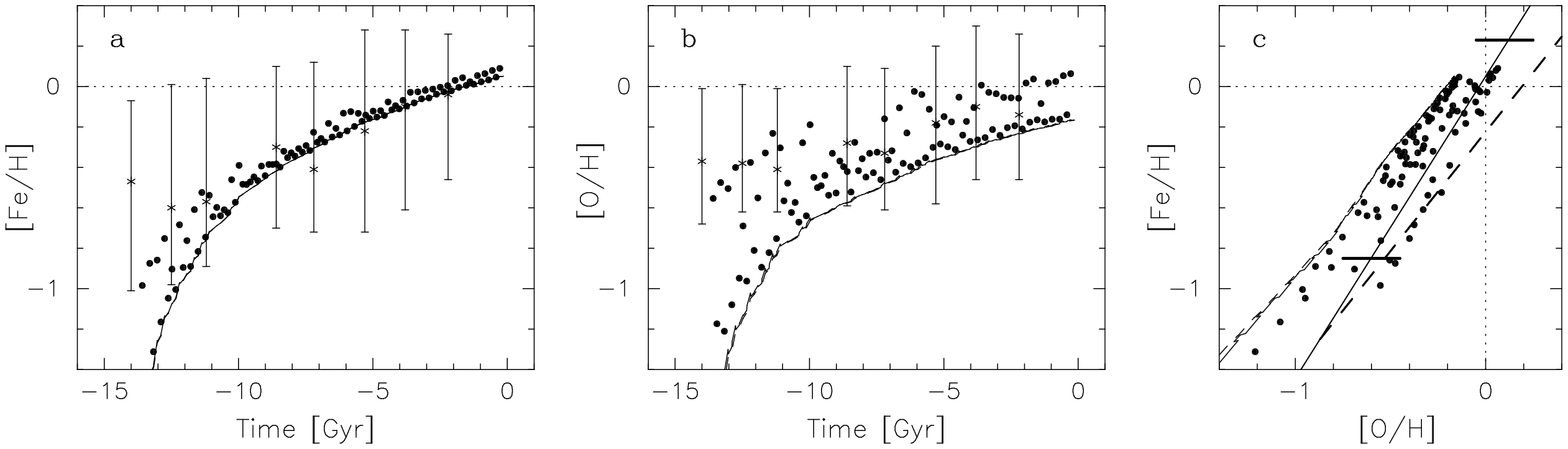,height=5.cm,width=16.cm,angle=270.}}
\vspace{-0.95cm} 
\end{figure*}
\begin{figure*}[ht]
\leftline{\psfig{figure=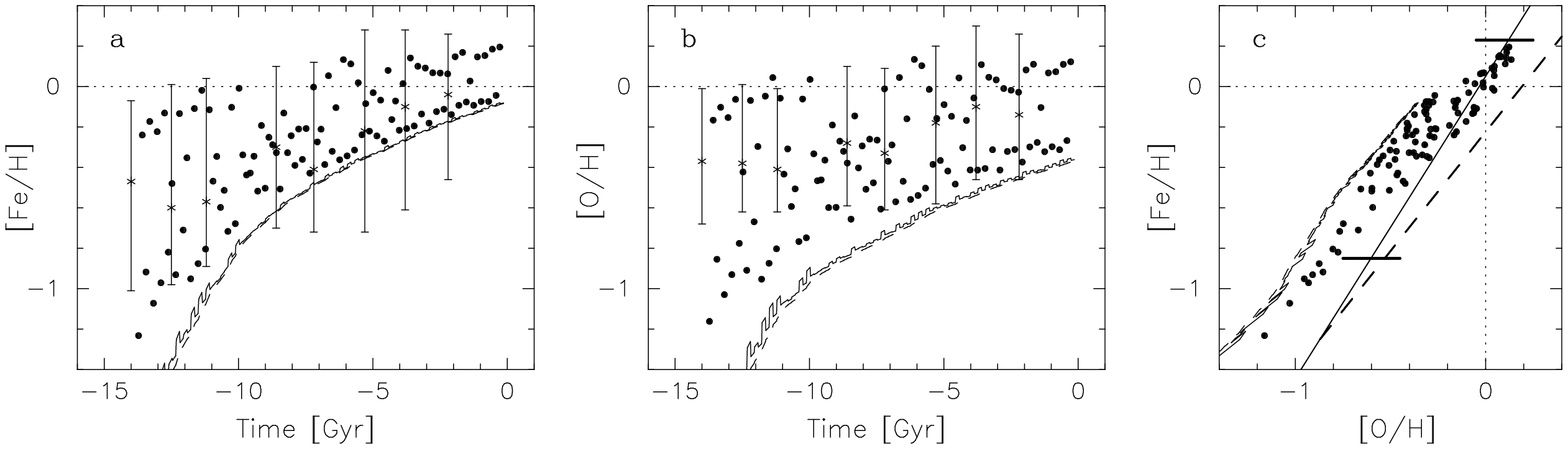,height=5.cm,width=16.cm,angle=270.}}
\vspace{-0.95cm} 
\end{figure*}
\begin{figure*}[ht]
\leftline{\psfig{figure=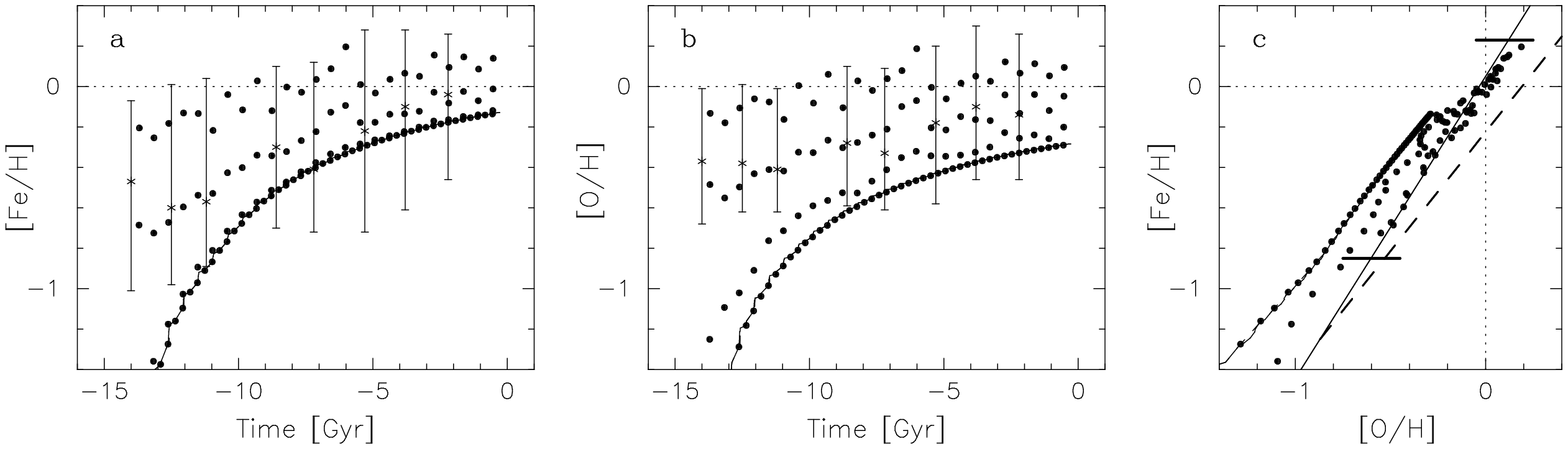,height=5.cm,width=16.cm,angle=270.}}
\vspace{-0.7cm} 
\caption[]{Model results for single and multiple sequential stellar
enrichment. {\em Top panels}: Single sequential enrichment assuming a
cloud dispersal time $t_{\rm disp}$ = 10$^{7}$ yr. {\em Center panels}:
Single sequential enrichment assuming $t_{\rm disp}$ = 2 10$^{7}$ yr.
{\em Bottom panels}: Multiple sequential enrichment (see text).
Model results are shown for variations of stellar and interstellar
[Fe/H] and [O/H] abundance ratios: {\bf a} [Fe/H] vs. age, {\bf b} [O/H] vs.
age,
{\bf c} [Fe/H] vs. [O/H]. Stellar abundances are indicated by filled circles.
Mean interstellar abundances (averaged over both active {\em and} inactive
clouds) are indicated by solid curves. Average interstellar abundances within
the inactive cloud ISM only (indicated by short dashed curve)
do approximately coincide with the overall mean abundances.
Remaining symbols and curves have the same meaning as in Fig. 4}
\end{figure*}

To maximize the effect of sequential enrichment on the stellar abundance 
variations, we assume $\vartheta = 0.2$, $\epsilon = 0.95$, 
$m_{\rm u}^{\rm SNII}$ = 25 \mss, $\phi^{\rm SNIa}$ = 0.005, and an 
enrichment efficiency $\lambda =0.95$ (\ie nearly all enriched stellar 
material ejected is mixed to the material wherein the {\em next} star 
formation event is induced). We consider cloud dispersal times 
$\Delta t_{\rm disp} \sim$ 10$^{7}$ yr. Such disperal times are 
among the largest ones deduced from observations of nearby star forming 
molecular clouds (see Sect. 3.3). Using a theoretical age vs. turnoff-mass 
relation (\eg Schaller \etal 1992), $\Delta t_{\rm disp}$ can be related to 
the least 
massive star $m_{\rm enr}$ able to enrich material before cloud dispersal. 
For instance, values of $\Delta t_{\rm disp} \sim$ 5 10$^{6}$, 10$^{7}$, and
2 10$^{7}$ yr, correspond to $m_{\rm enr} \sim$ 40, 15, and 12 \mss, 
respectively.

Figures 5-1 and 5-2 (top and center panels in Fig. 5, respectively) illustrate 
the effect of sequential enrichment on the stellar 
abundance variations for cloud dispersal times of $\Delta t_{\rm disp}$ = 
10$^{7}$ 
and 2 10$^{7}$ yr, respectively. 
The extent to which sequential enrichment contributes to the stellar 
abundance variations is determined by the chemical evolution of the ambient 
ISM. In general, large cloud dispersal times give 
rise to efficient locking up of metals in long living stars and  
enhanced stellar abundance variations relative to the abundances in the ISM.
Stellar abundance variations due to single sequential 
stellar enrichment are found to be maximal for $\Delta t_{\rm disp} 
\sim$2 10$^{7}$ 
yr. Larger values of $\Delta t_{\rm disp}$ allow stars less massive than 
$m \sim 12$ \ms to dilute the metal-rich ejecta of more massive stars
(\eg Hashimoto \etal 1993).

In case $\Delta t_{\rm disp}$ = 2 10$^{7}$ yr, resulting stellar abundance 
variations due to sequential enrichment are sufficiently large to explain 
the observed variations in [O/H]. In contrast, corresponding variations in 
[Fe/H] are much smaller than observed. This is true even though the models 
presented here do account for sequential enrichment by SNIb/c which 
usually show theoretical [O/Fe] ratios much lower than SNII (\eg Woosley 
\etal 1995. Results disagree with the observational fact that 
stellar abundance variations in [Fe/H] are considerably larger than in [O/H] 
(see also Gilmore \& Wyse 1991; Edvardsson \etal 1993a). 
An enhanced contribution of SNIb/c (\ie assuming $\phi_{\rm SNIb/c} > 0.33$)
or participation of SNIa to the process of sequential enrichment seems to be 
excluded by the observations (van den Bergh \& Tammann 1991; Tutukov \etal 
1992). 

We conclude that single sequential stellar enrichment models
are inconsistent with the observations because they: 1) result in [O/H] 
variations that are larger than those in [Fe/H],  2) predict current ISM 
abundances far below those observed, and 3) are difficult 
to reconcile with the apparent age independency of the stellar abundance 
variations (see Sect. 2). This conclusion is independent of the assumed 
cloud dispersion time scale $\Delta t_{\rm disp}$, sequential enrichment 
efficiency $\lambda$, value of $\vartheta$, background level of iron-group 
elements set by SNIa (\ie $\phi^{\rm SNIa}$), star formation efficiency 
$\epsilon$, and adopted IMF. Also, omitting the enrichment during the stellar 
wind phase of supernova progenitors does not alter this conclusion.

\subsubsection{Multiple sequential stellar enrichment}

The effect of sequential stellar enrichment on the abundance 
variations among successive stellar generations can be very large, especially 
for high sequential enrichment efficiencies and/or small amounts of 
cloud material to which the stellar ejecta are mixed before star formation is 
initiated. These conditions are naturally fullfilled when isolated gas clouds 
experience multiple star formation events (\ie $N_{\rm sf}^{\rm max}$$>$1)
before mixing with the surrounding 
ISM. Since the gas content of an isolated cloud is reduced by each star 
formation event, cloud abundances rapidly increase when enriched by 
successive generations of massive stars.

We consider two possible scenarios of multiple sequential enrichment.
In the first scenario, earlier generations of stars actually separate 
from the remaining subcloud material after dispersal of the subcloud 
core and do not further participate in the enrichment of the subcloud.
Such models result in substantial stellar abundance variations only under 
conditions and assumptions similar to those for the single sequential 
enrichment case. 
In the second scenario, all stellar generations formed in the subcloud 
continue to contribute to the enrichment until the entire subcloud has been 
dispersed. In this case, large stellar abundance variations arise due to
efficient recycling of the stellar ejecta from successive generations 
formed within the same cloud.

We apply the second scenario and consider all subclouds to experience 
multiple sequential enrichment. As maximum number of star formation 
events within one and the same subcloud we assume $N^{\rm max}_{\rm sf} = 4$ 
(\eg suggested by observations of OB subgroups in Orion; see Blaauw 1991).  
We adopt a sequential enrichment efficiency $\lambda = 1$ (by definition 
within the same subcloud), a star formation efficiency $\epsilon$ between 0.3 
and 0.6 for each star formation event, and $\Delta t_{\rm disp}$ = 10$^{7}$ yr.
Other values of these parameters may provide similar results.
Remaining model parameters are taken as for the single sequential enrichment 
model.

In Fig. 5-3, we plot resulting stellar abundance variations in case 
of multiple sequential stellar enrichment. Although variations 
caused by the first sequential enrichment event are relatively small (similar 
to the single sequential enrichment case assuming $\Delta t_{\rm disp} = 
10^{7}$ yr; 
see Fig. 5-1), abundance variations caused by subsequent events can be as large 
as $\sim$ 0.2$-$0.3 dex (depending on the subcloud abundances). As 
mentioned before,  
such large abundance variations are mainly due to ongoing sequential 
enrichment of the remaining cloud material by stellar generations formed 
earlier in the cloud. We find that models incorporating 
multiple sequential stellar enrichment encounter the same problems as single
sequential enrichment models. However, the former models appear 
observationally far more justified. This is true in particular for the
sequential enrichment and star formation efficiencies, as well as the cloud 
core dispersal times, required to obtain a given stellar abundance variation.

We conclude that the observed stellar abundance variations are difficult to 
explain by sequential stellar enrichment only. 
This conclusion is not altered when allowing for variations in sequential 
enrichment between distinct star formation events (\eg by considering 
variations in the IMF, relative formation rates of SNII and SNIb/c, etc.).
Possible exceptions may be selective mixing of SNII nucleo-synthesis products 
to the material wherein star formation is induced, and/or cloud 
conditions that determine both the composition of the ejecta returned by
a stellar generation (\eg by means of the IMF, $m_{\rm u}^{\rm SNII}$, 
$\Delta t_{\rm disp}$, contribution by SNIb/c, binaries) and the sequential 
enrichment efficiency (and/or star formation efficiency). However, the impact 
of such conditions on the stellar abundance variations, which would imply that
the IMF weighted SN yields used here would be considerably modified, is 
beyond the scope of this paper.

\subsection{Episodic infall of metal-deficient matter}

\begin{figure*}[ht]
\leftline{\psfig{figure=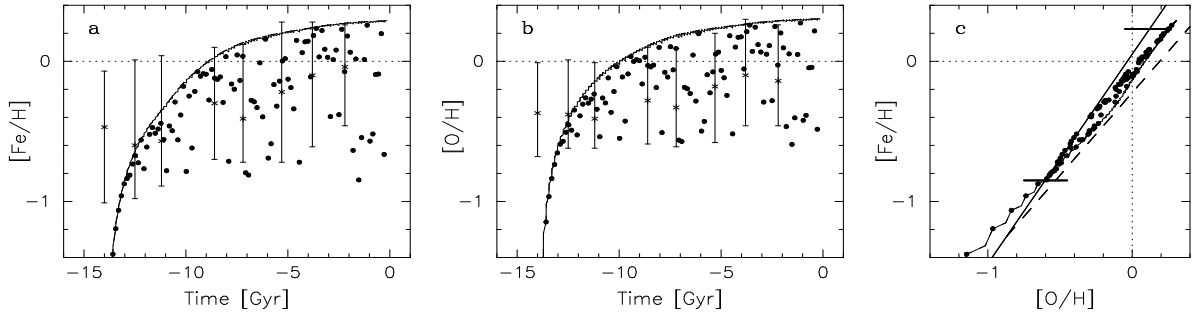,height=5.cm,width=16.cm,angle=270.}}
\vspace{-0.7cm} 
\caption[]{Infall of metal-deficient material:
model results are shown for variations of stellar and interstellar [Fe/H]
and [O/H] abundance ratios: {\bf a} [Fe/H] vs. age, {\bf b} [O/H] vs. age,
{\bf c} [Fe/H] vs. [O/H]. Symbols and curves have the same meaning as in Fig. 5}
\end{figure*}

Infall of metal-poor material can account for abundances of newly formed stars 
which lie substantially below the abundances in the global disk ISM. 
In principle, abundance variations due to metal-deficient gas infall are
determined by the abundances within the infalling gas, the gas infall rate, 
and the amount of disk ISM to which the infalling material is mixed before star 
formation is initiated. Stellar abundance variations due to infall of 
metal-rich material associated with SN ejecta from massive stars in the 
Galactic disk can be considered as a special case of sequential stellar 
enrichment and is not discussed here. 

Element abundances within the infalling gas are constrained by the lowest 
abundances observed for disk stars with [Fe/H]$\ga -1$. 
The Edvardsson \etal data imply infall abundances of [M/H]$_{\rm inf} \la 
-$0.8 to $-$1.2 for \eg M=C, Mg, Al, and Si. These abundances are consistent 
with observations of interstellar clouds in the halo (see Sect. 5.2) and 
suggest that infall induced star formation is associated with 
the lowest abundances observed among disk F and G dwarfs in the SNBH.
We assume infall abundances similar to the abundances observed among the oldest 
metal-poor disk stars, \ie [Fe/H] = $-$1, [O/H] = $-$0.65, and a hydrogen mass 
fraction of X$\sim$0.72 (\eg Bessell \etal 1991). 
For simplicity, we do not account for abundance inhomogeneities within the 
infalling gas and assume infall abundances to be constant in time.
Furthermore, we consider infall to occur as soon as the infall abundances are 
reached in the global disk ISM (presumably corresponding with the onset of star 
formation in the disk).

The detailed manner in which gas infall varies with time is not 
essential for the results presented here {\em as long as} infalling gas 
plays an important role in determining the stellar abundances when it 
induces star formation. 
Here, we deal with the concept of infall induced star 
formation, \ie the infalling gas initiates star formation when falling onto the 
disk. This concept is based on observations of infalling high-velocity 
clouds that strongly interact with disk ISM and initiate star formation therein
as soon the critical density for star formation is reached (see Sect. 5.2). 
Since star formation occurs by definition within active subclouds according 
to our model, infall is associated with subclouds experiencing star 
formation shortly after their formation. For simplicity, we assume each 
subcloud to contain an amount of infalling gas accumulated at the time star 
formation is induced. This amount is taken as a random fraction of the 
initial subcloud mass (\ie between 0 and 1). 
In this manner, we allow for local and episodic gas infall onto
the Galactic disk ISM. On average, the gas infall 
rate is assumed to decay exponentially on a time scale $t_{\rm dec} = 6$ Gyr 
while its overall amplitude is constrained by $\mu_{1}$ = 0.05$-$0.2
(\cf Eq. (1); see Table 3). 

We assume an initial disk mass $M_{\rm cl}$ = 3 10$^{10}$ \ms before the 
onset of gas infall. This results in a disk initial-to-final mass-ratio 
$\zeta =$0.55 according to an exponentional decrease of subcloud mass with 
cloud evolution time ($M_{\rm scl}(0) =$ 2 10$^{9}$ \mss; \cf Table 3).
Clearly, stellar abundance variations due to metal-deficient gas infall
are small at low levels of enrichment of the global disk ISM (\ie large 
initial disk mass). Furthermore, continuous and large scale metal-deficient gas
infall not associated with star formation results in relatively small stellar 
abundance variations. Thus, the magnitude of the stellar abundance 
variations is affected by the ratio of initial disk mass and total 
amount of infalling matter, as well as the amount of infalling gas that is 
involved with induced star formation in the disk ISM.

Figure 6 displays resulting AMRs for iron and oxygen in case of episodic infall 
of metal-deficient gas. We assumed $\phi^{\rm SNIa}$ = 0.015, 
$\phi^{\rm SNIb/c}$ = 0.33, and $m_{\rm u}^{\rm SNII}$ = 40 \mss. The value 
of $m_{\rm u}^{\rm SNII}$ is taken larger than for the reference model (\cf 
Table 2) to obtain somewhat better agreement with the observations.
Resulting abundances in the disk ISM follow the upper 
end of the abundances observed in F and G dwarfs younger than $\sim$10 Gyr.
Although the effect of global infall of metal deficient gas is generally 
to dilute the enrichment of the ISM, the inflow model results in larger ISM 
abundances than the reference model. This is mainly due to: 1) the 
asumption of a low initial disk mass which allows for a rapid early 
enrichment of the disk, and 2) the assumption of {\em local} infall of 
metal-deficient gas and subsequent star formation therein so that infalling 
material has relatively small effect on the dilution of the {\em global} 
disk ISM.

By varying the ratio of infalling matter and disk ISM within each 
subcloud, stellar abundance variations of $\Delta$ [Fe/H] $\sim$ 0.8 dex and 
$\Delta$ [O/H] $\sim$ 0.65 dex naturally can be accounted for. 
In addition, the scatter in [Fe/H] remains larger than in [O/H] 
since the iron abundances within the infalling gas relative to solar 
(\ie [Fe/H]$_{\rm inf}$ = $-$1) are much smaller than that of oxygen 
([O/H]$_{\rm inf}$ = $-$0.65). We note that the current gas infall 
rate of $\sim$3.1 \ms yr$^{-1}$ predicted by the model shown in 
Fig. 6 is larger than suggested by the observations (see Sect. 5.1). 
However, other choices of model parameters, \eg $\zeta$, predict 
much lower infall rates while providing similar abundance results.

Our models incorporating metal-deficient gas infall are in good agreement 
with the observed magnitude of stellar abundance variations and the 
slope of the [Fe/H] vs. [O/H] relation observed. However, these models 
predict current interstellar [Fe/H] and [O/H] abundance ratios of 
$\sim$0.2 dex {\em above} solar. This is in marked contrast with 
[O/H] abundance ratios of $\sim 0.15$ dex {\em below} solar observed both in 
interstellar gas and recently formed stars in the SNBH (see Sect. 2).
In addition, these models appear to disagree with the observations 
on two other grounds. First, no significant scatter in the [Fe/H] vs. 
[O/H] relation is predicted, contrary to what is observed for intermediate 
age disk stars (\cf Fig. 1). Part of the observed scatter may be due to 
experimental errors but variations of at least $\pm$0.1 dex in the [Fe/H] vs. 
[O/H] relation are probably real and have to be explained by any satisfactory 
model. A way to account for such scatter would be
to allow for considerable abundance variations among different parcels of 
infalling gas. Secondly, these models predict stellar abundance variations 
to increase with time. This is inconsistent with the apparent constancy of 
the abundance scatter observed. Possible ways out may be uncertainties
in the ages of stars older than $\sim$ 10 Gyr or disk evolution times 
in excess of $t_{\rm ev} \sim 14$ Gyr. 

We conclude that models dealing with metal-deficient gas infall can 
probably be excluded as the complete answer to the stellar abundance variations 
observed, even though such models are in good agreement with both the observed 
abundance variations and [Fe/H] vs. [O/H] relation.
This conclusion is primarily based on the fact that such models predict mean 
current ISM abundances $\sim$0.4 dex larger than those observed in the SNBH.

\subsection{Metal-poor gas infall combined with sequential enrichment}

Motivated by the results previously discussed, we study the combined 
effect of sequential stellar enrichment and episodic infall of metal-deficient 
gas on the inhomogeneous chemical evolution of the Galactic disk. 
Such investigation is important also because these processes are observed to 
operate simultaneously in the SNBH (see Sect. 5.2).

\begin{figure*}[htp]
\leftline{\psfig{figure=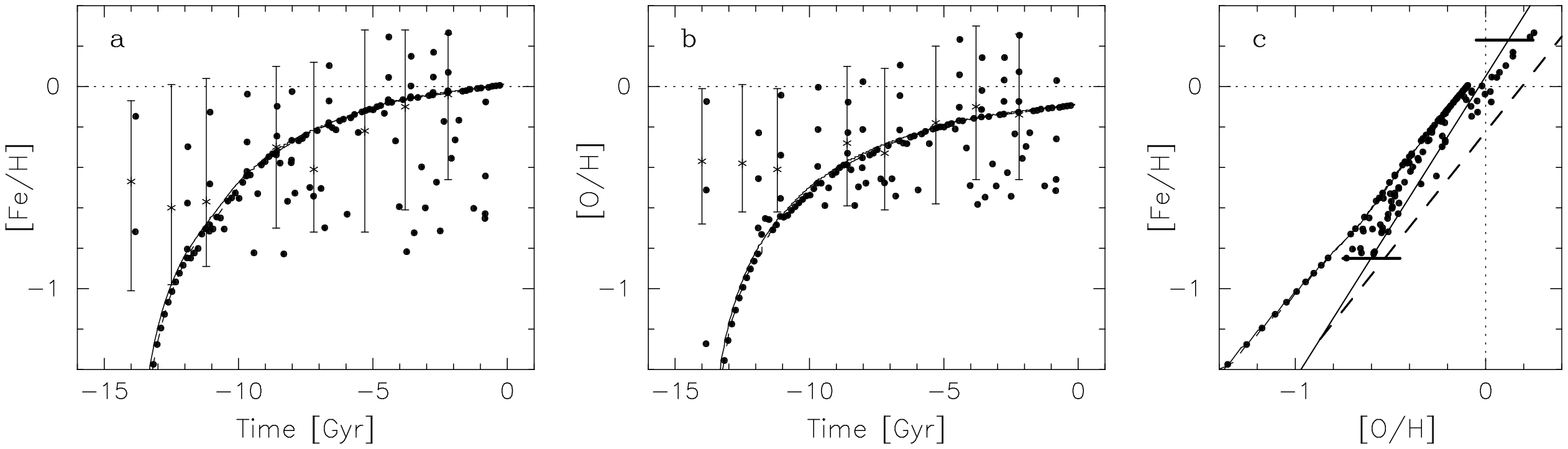,height=5.cm,width=16.cm,angle=270.}}
\vspace{-0.8cm} 
\end{figure*}
\begin{figure*}[ht]
\leftline{\psfig{figure=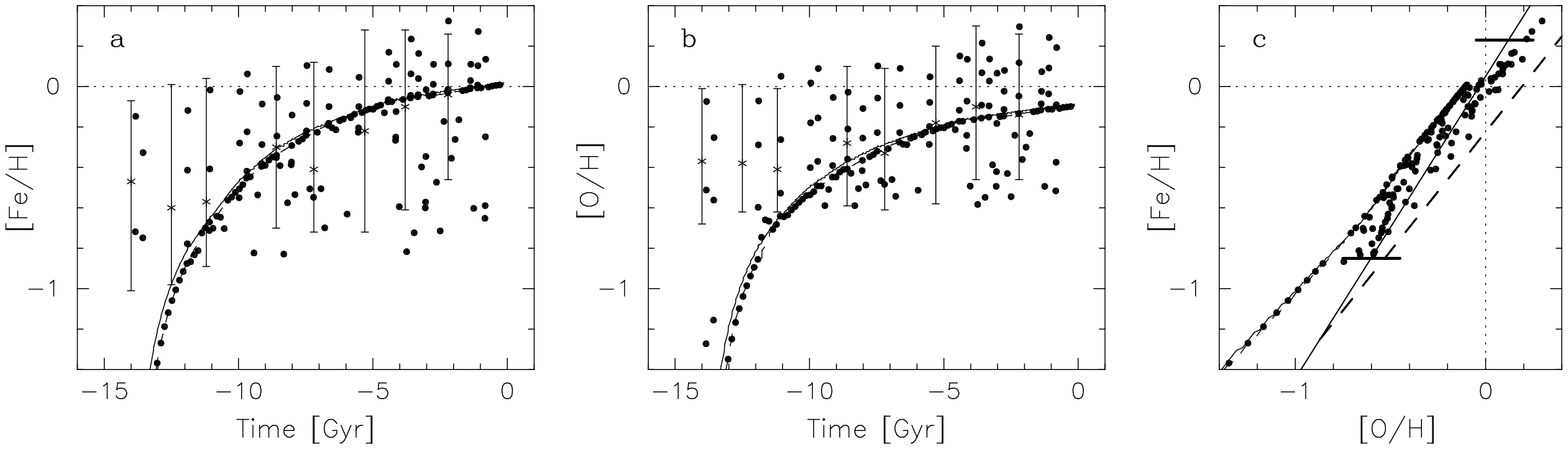,height=5.cm,width=16.cm,angle=270.}}
\vspace{-0.8cm} 
\end{figure*}
\begin{figure*}[ht]
\leftline{\psfig{figure=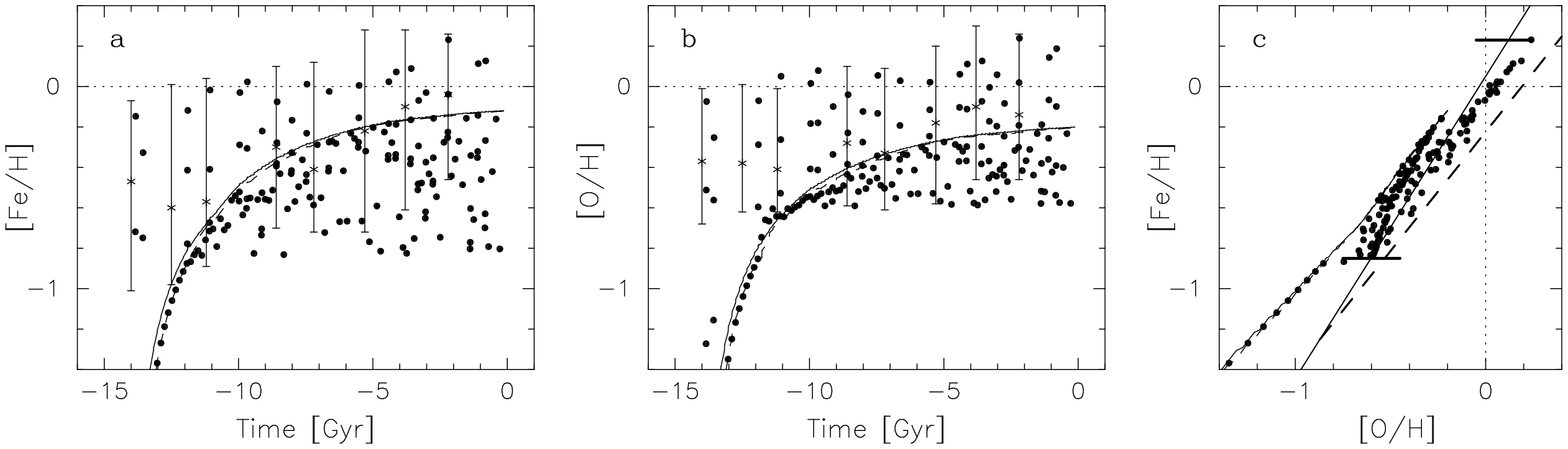,height=5.cm,width=16.cm,angle=270.}}
\vspace{-0.7cm} 
\caption[]{Model results for combined sequential stellar
enrichment and episodic infall of metal-poor gas.
{\em Top panels}: $\sim$10\% of the clouds is assumed to experience multiple
sequential enrichment while half of the subclouds is involved with
metal-deficient gas infall.
Nearly 5\% of the clouds undergoes {\em both} infall of metal-deficient
material and sequential stellar enrichment.
{\em Center panels}: 25\% of the subclouds experiences multiple
sequential enrichment while half of the subclouds is involved with metal-poor
gas infall.
{\em Bottom panels}: as center panels but
{\em all} subclouds experience metal-deficient gas infall.
Model results are shown for variations of stellar and
interstellar [Fe/H] and [O/H] abundance ratios: {\bf a} [Fe/H] vs. age,
{\bf b} [O/H] vs. age, {\bf c} [Fe/H] vs. [O/H]. Symbols and curves have the
same meaning as in Fig. 5}
\end{figure*}

Attractive features of combined infall of metal-poor gas and sequential 
stellar enrichment are that a self-consistent explanation can be obtained for: 
1) the presence of high metallicity stars at early epochs of star formation 
in the Galactic disk (due to sequential enrichment), 2) the presence of 
metal-poor stars at recent epochs of Galactic evolution (as a result of 
metal-deficient gas infall), 
3) the nearly constant magnitude of the stellar abundance variations 
during the lifetime of the disk, and 4) abundances in the local disk ISM
that are currently below solar (as observed for oxygen). 

We show in Fig. 7 results for combined sequential stellar enrichment and 
metal-deficient gas infall. Model assumptions concerning each of these 
processes are similar to those described in the previous sections 
(\eg $\epsilon^{\rm max} = 0.90$, 
$\lambda = 0.95$, $m_{\rm u}^{\rm SNII}$ = 25 \mss, 
$\phi^{\rm SNIa} = 0.005$, and $\phi^{\rm SNIb/c} = 0.33$; \cf Tables 2 and 3).
The three models shown in Fig. 7 differ only in the amounts of disk ISM 
involved with sequential stellar enrichment and infalling gas accreted during 
the lifetime of the disk. 
For each of these models, resulting stellar abundance variations and [Fe/H] vs.
[O/H] relation are consistent with the observations.
Clearly, models with combined sequential enrichment and metal-poor gas infall 
do not encounter the specific problems involved when each of these processes 
is considered separately. 

We study the relative impact of metal-poor gas infall and sequential 
stellar enrichment on the resulting stellar abundance variations as well as 
the global enrichment of the ISM. First, we investigate the effect of varying 
the fraction of subclouds (\ie the amount of 
star forming disk ISM) experiencing multiple 
sequential stellar enrichment ($N^{\rm max}_{\rm sf}$=4, 
$\Delta t_{\rm disp}$ = 10$^{7}$ yr, $\epsilon^{\rm max}$ = 0.9; 
see Sect. 4.2). This fraction increases from $\sim$10\% to $\sim$25\%
when going from top to center models shown in Fig. 7. The remaining part of 
the subclouds is assumed to experience one single star formation event. 
Furthermore, we assume half of the subclouds to form stars partly from infall 
of metal-deficient gas (\ie Figs. 7-1 and 7-2), regardless of the
number of star formation events in each subcloud. 
Note that subclouds involved with metal-poor gas 
infall form predominantly stars with abundances below those in the global disk 
ISM.
It can be seen that mean interstellar abundances and stellar abundance 
variations are not significantly altered when the fraction of ISM associated 
with sequential stellar enrichment is increased from 10 to 25\%. 
However, when this fraction is further increased, more and more metals 
will be locked up in long living stars due to sequential enrichment and
marked deviations from the observed [Fe/H] vs. [O/H] relation will occur
(see Sect. 4.2).

Secondly, we investigate the effect when the fraction of subclouds 
forming stars from metal-deficient gas infall is increased from 50 to 100\%. 
In this case, stellar generations are all formed according to infall induced 
sequential star formation and the total mass of infalling gas is increased 
by a factor two. This results in a reduction of the interstellar [Fe/H] and 
[O/H] abundance ratios by $\sim$0.1 dex (see Fig. 7-3). 
Interestingly, this marginally affects the magnitude of the resulting stellar 
abundance variations (assuming $\epsilon^{\rm max} =0.9$, $\lambda = 0.95$) 
but strongly alters the number of stars with abundances below those present in 
the global disk ISM. We note that direct comparison of the abundance results 
with previous models is not justified because the enhanced gas infall 
model results 
in a current gas-to-total mass-ratio $\mu_{1} \sim$0.3, \ie considerably higher 
than the $\mu_{1} =$ 0.1$-$0.2 indicated by the observations (\cf Table 3).
To arrive at $\mu_{1} = 0.2$, a reduction in initial disk mass from 
3 10$^{10}$ to 2 10$^{10}$ \ms would be required. In turn, this would result 
in ISM abundances and stellar abundance variations similar to that for 
models with more modest infall rates. 

\subsection{Additional abundance constraints}

Keeping these results in mind, we study how models with combined 
sequential stellar enrichment and metal-poor gas infall behave when confronted 
with additional observational constraints provided by the stellar abundance 
variations and current ISM abundances of C, Mg, Al, and Si. 

Carbon abundance data for 85 F and G dwarfs in the SNBH have been 
presented by Andersson \& Edvardsson (1994). These data show that there is 
a weak correlation between [C/H] and [O/H] (see Fig. 8). The shape of this 
correlation differs from that between \eg [Fe/H] and [O/H].
In addition, the variation in [C/H] (\ie $\ga$0.6 dex) at a given value of 
[O/H] is about three times larger than that in [Fe/H].

If the observed stellar abundance variations are caused by infall of 
metal-deficient gas only, one would expect that stellar abundances for all
elements heavier than helium would be mutually correlated, \eg similar to the 
correlation between oxgyen and iron. In case of sequential stellar enrichment 
only, a similar behaviour would be expected only for elements that are produced 
predominantly by SNII and SNIb/c. 
This implies that abundance-abundance variations between elements which are 
not synthesized predominantly within SNII and SNIb/c
(such as C and N), on the one hand, and elements that are produced 
predominantly within supernovae (\eg O, Si), on the other hand, may be 
conclusive about the importance of metal-deficient gas infall. 

We show in Fig. 8 results for: 
1) infall of metal-deficient gas only (\cf Fig. 6), 2)
multiple sequential stellar enrichment only (\cf Fig. 5-1), and 3) combined
metal-poor gas infall and sequential enrichment (\cf Fig. 7-1).
The infall model predicts no substantial scatter in the [C/H] vs. [O/H] 
relation but follows the trend in the observations well. 
Although the scatter in the [C/H] vs. [O/H] relation suggests that infall is 
not exclusively responsible for the observed variations, the shape of 
this relation indicates that infall is important.
Conversely, the sequential enrichment model shows large scatter in the [C/H] 
vs. [O/H] relation but appears to deviate from the observed trend.
The correlation predicted by the combined sequential enrichment and infall 
model appears in reasonable agreement with the observations.
However, the observed carbon abundances at [C/H]$\sim$0. exhibit considerable 
more scatter than predicted by the model shown in Fig. 7-1 and seem to 
require a somewhat steeper 
increase of carbon relative to oxygen. This may be due to variations in 
sequential stellar enrichment between different star formation events and/or 
variations in abundances within the infalling material.

\begin{figure*}[htp]
\leftline{\psfig{figure=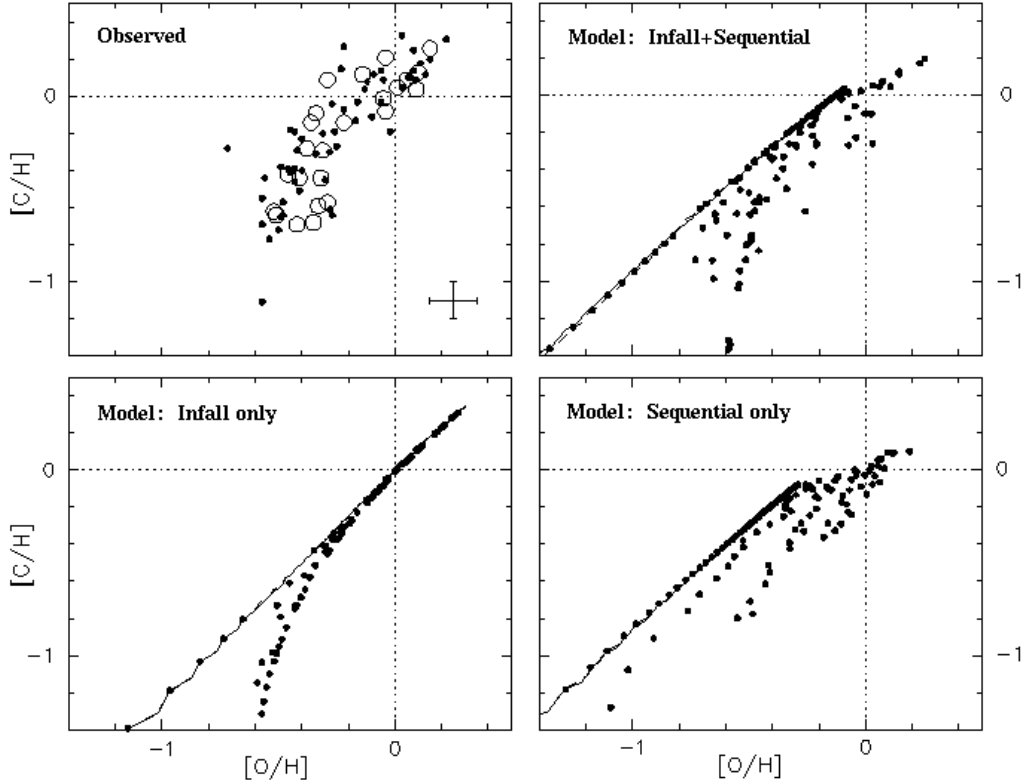,height=12.cm,width=17.cm,angle=270.}}
\vspace{-0.7cm} 
\caption[]{Comparison between observed and model-predicted
[C/H] vs. [O/H] relation. {\em Observations}: data for F and G
dwarfs in the SNBH from Andersson \& Edvardsson (1995).
Open circles represent stars with mean stellar galactocentric
distances at birth within 0.5 kpc from the Sun ($R_{\odot}$ = 8.4 kpc).
Full dots indicate stars with average distances within $\sim$2 kpc from the Sun.
Typical errors are indicated at the bottom right (top left panel).
{\em Model results}: Predicted abundances of stars (full dots) and gas
(solid line) for models incorporating metal-poor gas infall and/or sequential
stellar enrichment}
\end{figure*}

In our models, the overall shape of the stellar [C/H] vs. [O/H] relation 
is due to infall of metal-poor material with carbon abundances 
[C/O]$_{\rm inf} \approx -$0.4. The reason why these relatively low carbon 
infall abundances are necessary to explain the observed 
trend in the [C/H] vs. [O/H] relation, is unclear. A possible explanation 
may be a delayed carbon enrichment of the disk ISM, \eg by low-mass SNIa 
progenitors that experience incomplete carbon burning or by low-mass AGB stars 
with small carbon yields. 

Sequential stellar enrichment seems inevitable to explain the observed scatter 
in the [C/H] vs. [O/H] relation. Although large abundance inhomogeneities 
within 
the infalling gas may reproduce the observed variations as well, such 
inhomogeneities in [C/O] appear inconsistent with the small scatter observed 
in \eg [Fe/O]. Also, uncertainties in the derived carbon abundances may be
considerably larger than those in O and Fe but are not likely to exceed 
$\sim$0.2 dex (see Andersson \& Edvardsson 1994; see also EDV). Therefore, it 
seems improbable that the scatter observed in the [C/H] vs [O/H] relation is 
due to observational errors. Finally, it is difficult to see how chemical 
differentiation processes (\eg dust depletion, element mixing to the 
surface, O/N-cycle, metallicity dependent nucleo-synthesis, etc.) can cause 
such large abundance-abundance variations among stars similar in mass and age. 

\begin{figure*}[htp]
\leftline{\psfig{figure=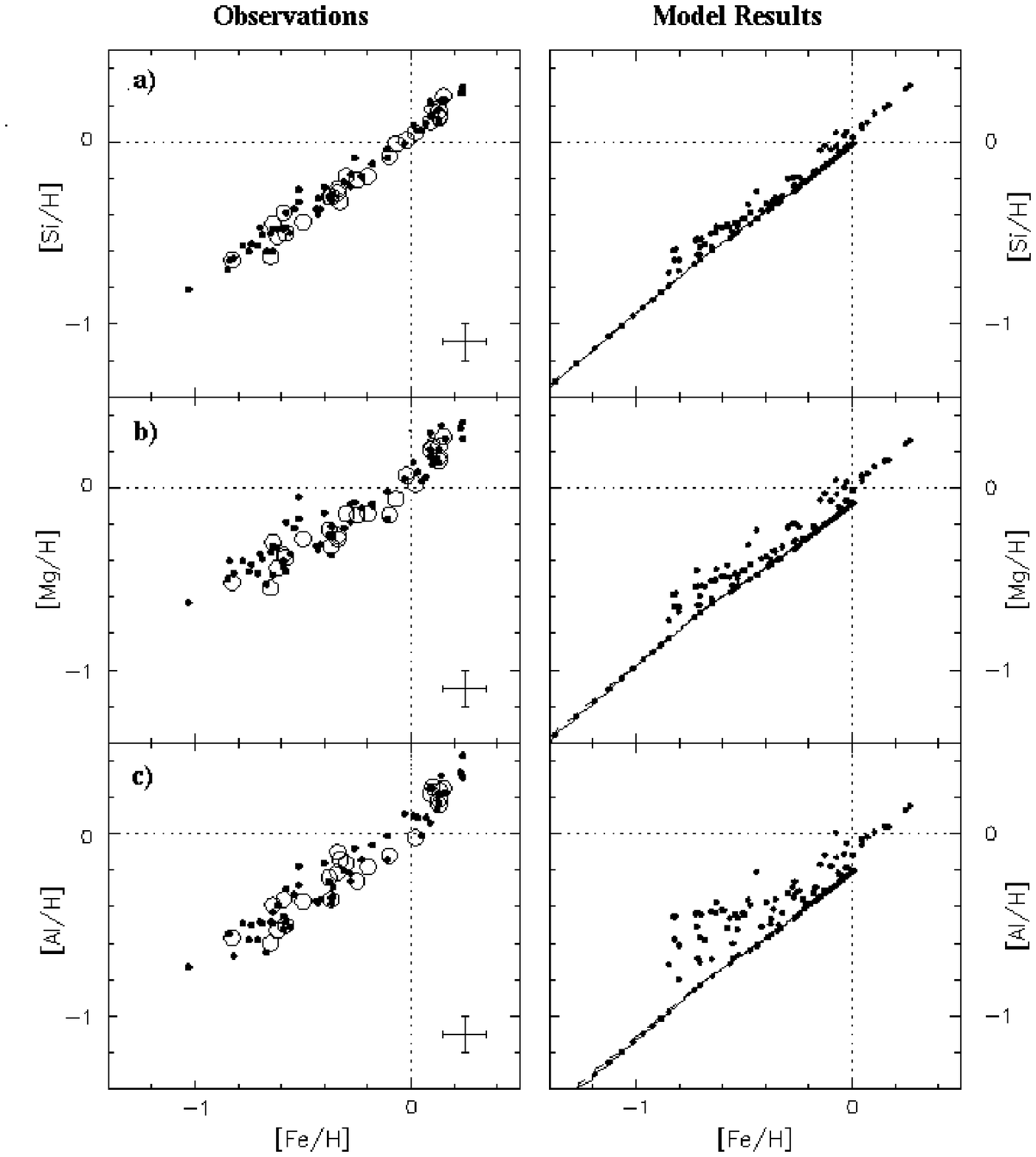,height=18.cm,width=16.cm,angle=0.}}
\vspace{-1.1cm} 
\caption[]{Comparison between observed (left) and model-predicted
(right) [M/H] vs. [Fe/H] relations: a) [Si/Fe], b) [Mg/Fe], and c) [Al/Fe].
{\em Observations}: data for F and G dwarfs in the SNBH from Edvardsson \etal
(1993a). Symbols have the same meaning as in Fig. 8. Typical errors are
indicated at the lower right of each panel. {\em Model results}: predicted
abundances of stars (full dots) and gas (solid line) for model 7-1}
\end{figure*}

From the arguments above, we conclude that models incorporating {\em both} 
infall of metal-poor gas and sequential stellar enrichment provide an
adequate explanation for the observed [C/H] vs. [O/H] relation. 
For the model shown in Fig. 7-1, we compare the predicted stellar abundance 
ratios Si, Mg, and Al vs. Fe with the observations in Fig. 9. 
We find that slight offsets in [M/H] are present between the model predicted 
and observed relations. These offsets, most pronounced for Al, are due to 
details in the adopted stellar yields and related parameters (see Table 3), 
and are not essential for the following discussion. Interestingly, a number 
of observed features are naturally reproduced by these models.

First, the observations 
suggest that variations in the stellar [M/H] abundance ratios decrease 
with increasing metallicity. This is theoretically predicted by the 
individual effects of both sequential stellar enrichment and infall of 
metal-deficient material (as discussed above). For the model shown in Fig. 
7-1, the resulting scatter in 
[M/H] at a given value of \eg [Fe/H] is mainly due to sequential enrichment 
(except at abundances [M/H]$\la -0.7$) and is strongly related 
to the iron contribution by SNIa in regions that do not experience sequential 
enrichment. It can be seen that the scatter in [M/H] strongly 
decreases at solar metallicities and above. This may indicate that 
much of the observed element-to-element variations at high metallicities is 
due to observational errors of $\sim$0.1 dex in [M/H]. Alternatively, the 
observed scatter at high metallicities may imply that sequential enrichment
by massive stars varies from one star formation event to another, \eg by 
means of variations in the upper mass limit for SNII. 

Secondly, the predicted variation in [Mg/H] at a given [Fe/H] is substantially 
larger than that in [Si/H], consistent with the observations. In our model,
this is due to: 1) the fact that part of the Si comes from SNIa which 
are important contributors also to Fe (Mg is produced less efficiently 
in SNIa than is Si by about one order of magnitude; \eg Nomoto \etal 1984), 
and 2) the predicted ISM abundance of [Mg/H] is less by about 0.1 dex than 
that of [Si/H] (see Fig. 9).

Thirdly, the observed variation in [Al/H] is similar 
(or even larger) than that in [Mg/H] (see EDV). This behaviour is also 
found for the model shown in Fig. 7-1. The predicted ISM abundance of Al is 
probably 
too low by $\sim$0.1 dex so that the resulting abundance scatter would be 
slightly reduced when correcting for this.
The impact of sequential stellar enrichment is more 
pronounced for Al than for Mg and Si, due to the somewhat lower ISM 
abundance of Al (even after correction). Overall, we conclude that the 
magnitudes of the resulting stellar variations in Si, Mg, and Al vs. Fe
appear in reasonable agreement with the observations.

Comparison of variations in Mg, Si, and Al vs. O with the observations reveals 
a somewhat different picture from that vs. Fe. The observed variation of 
[Mg/H] with [O/H] has been shown in Fig. 1; Al and Si display a similar
behaviour.
No trend is observed for variations in the scatter in the $\Delta$[M/H] vs. 
[O/H] relation, in contrast to that in the $\Delta$[M/H] vs. [Fe/H] relation. 
Furthermore, the observations indicate mean variations in [M/Fe]
{\em of the same magnitude} as those in [M/O].
In contrast, the model shown in Fig. 7-1 predicts variations in Mg, Al, and 
Si vs. Fe that 
are {\em considerably larger} than those vs. O. This is simply due to the
fact that a substantial fraction of Fe originates from SNIa. 
Therefore, models predict hardly any scatter in \eg [Mg/O] 
since both elements are synthesized predominantly within SNII (and SNIb/c). 
This implies that either the scatter in the observed [Mg/H] vs. [O/H] relation 
is due to observational errors or that an additional process is needed in 
the models to explain this scatter.

In the former case, there would be no reason to believe the variations observed 
in the [M/H] vs. [Fe/H] relations either. However, we have argued above 
that at least part of this scatter is real. 
In the latter case, an additional mechanism causing the scatter in 
[M/O] could be variations from one star formation site to another
in the enrichment by SNII (and SNIb/c). Alternatively, such variations  
could include variations in \eg the IMF, upper mass limit for SNII, 
and/or the mass distribution of binaries. 
We have verified that such variations generally result in 
abundance-abundance scatter sufficiently large to account for the observed 
variations of 0.1 dex in [M/O] and sufficiently small to have a negligble 
effect on the scatter observed in [M/Fe].
Clearly, element-to-element variations in enrichment from one star 
formation site to another would be a natural refinement of 
the sequential stellar enrichment models discussed before.

We conclude that models incorporating both sequential 
stellar enrichment and episodic infall of metal-poor gas provide a natural 
explanation for the observed stellar abundance variations and are consistent 
with the ISM abundances of C, O, Fe, Mg, Si, and Al. 
We find that the mean ISM abundances and abundance-abundance relations can 
provide only limited constraints on the relative importance of sequential 
enrichment and infall induced star formation in the Galactic disk.
Therefore, improvements in observational and 
theoretical constraints are required to disentangle the effects of these 
processes on the inhomogeneous chemical evolution of the Galactic disk 
in a more quantitative way.

\section{Discussion}

We briefly examine how the combined sequential stellar enrichment and 
metal-deficient gas infall models discussed in the previous section behave 
when confronted with independent constraints provided by the current star 
formation rate in the Galactic disk and the chemical evolution of the 
Galactic halo. Thereafter, we discuss observational evidence in support of 
sequential enrichment and gas infall in the local disk ISM and consider 
possible implications of these processes for the chemical evolution of the 
Galaxy as a whole.

\subsection{Additional constraints}

\subsubsection{SFR related constraints}

The combined sequential enrichment and metal-deficient gas infall (Fig. 7-1) 
predicts a present SFR of $\sim$5.2 \ms yr$^{-1}$, and current 
rates of SNII (excluding SNIb/c) and SNIa of $R^{\rm SNII}$ = 2.1 10$^{-2}$ 
yr$^{-1}$ and $R^{\rm SNIa}$ = 1.3 10$^{-3}$ yr$^{-1}$, respectively.
These values are roughly consistent with the observations (\ie within a 
factor of two; see Table 3). Adopted values of $m_{\rm u}^{\rm SNII}$=25 
\ms and $\psi^{\rm SNIa}$=0.005 may be somewhat too low since the SN-rates 
scale with the predicted SFR.
For the same model, the current gas infall rate is determined by the assumed 
disk initial-to-final mass-ratio $\zeta$=0.5 and by the time scale 
$\tau_{\rm inf}$= 6 Gyr on which infall decays exponentially. This results in 
a current gas infall rate of 1.8 \ms yr$^{-1}$. Observations indicate a
current gas infall rate of $\sim$0.5 \ms yr$^{-1}$ 
(\eg Mirabel \& Morras 1984). However, selection effects may account for an 
underestimate of a factor of 2$-$3 (see Sect. 5.2).
We note that higher values of $\zeta$ and/or lower 
values of $\tau_{\rm inf}$ may apply equally well. 

The predicted rates above all scale with the amplitude of the SFR. In turn, 
this 
amplitude is determined by the total cloud mass M$_{\rm cl}=$2 10$^{11}$ \ms 
and SFR decay time scale $t_{\rm decr} \sim 6$ Gyr assumed (see Sect. 3.2). 
Distinct values of M$_{\rm cl}$ and/or $t_{\rm decr}$ will not affect
the predicted stellar and interstellar abundances substantially, provided that 
a current gas-to-total mass-ratio $\mu_{1} = 0.1$ is maintained.
We conclude that the adopted parameters for the model shown in Fig. 7-1 are 
consistent with observational constraints on the current SFR, gas infall rate, 
and supernova rates. Obviously, these observations do not yet provide tight 
constraints on \eg $t_{\rm decr}$, $\mu_{1}$, and $\zeta$, thus preventing a 
clear distinction between chemical evolution models based on these quantities
(\cf Table 3).

\subsubsection{Constraints related to the enrichment of the Galactic halo}

The mean plateau value of [O/Fe]$\sim$0.5 $\pm$0.15 observed for halo 
stars with [Fe/H]$\la -1$ (\eg Bessell \etal 1991) is presumably determined 
by the average [O/Fe] ratio within the ejecta of SNII (and SNIb/c) as well 
as the initial abundances within the halo ISM. 
For our models, the plateau value 
implies a maximum upper mass limit of SNII progenitors of $m_{\rm u}^{\rm SNII} 
\approx$40 \mss, assuming initial metallicities [Fe/H]$\la-$1, a Salpeter 
IMF, and stellar yields as described in Sect. 3.4. An even larger value for 
$m_{\rm u}^{\rm SNII}$ is implied when SNIb/c contributed substantially to the 
halo enrichment.

We assumed $m_{\rm u}^{\rm SNII}$ = 25 \ms for the combined infall + 
sequential enrichment model (Fig. 7-1) discussed before. This results in
[O/Fe]$\sim$0.2 at [Fe/H]$\la -1$ while omitting the contribution 
from SNIb/c would have resulted in [O/Fe]$\sim$0.25 at [Fe/H]$\la -1$. 
This is inconsistent with the observations. 
Possible solutions to this discrepancy are: 
1) $m_{\rm u}^{\rm SNII}$ and/or the IMF have changed between the time  
stars formed in the halo and the time of onset of star formation in the disk, 
2) $m_{\rm u}^{\rm SNII}$ is actually $\sim$40 \ms for 
disk stars so that the predicted current disk ISM abundances of \eg O and Fe 
increase and the effect of sequential enrichment is reduced, and/or 
3) the adopted yields for SNII (and or SNIb/c) are in error at metallicities 
below [Fe/H]$\sim -1$.

Although the first two possibilities cannot be excluded, we favour the latter 
option since values of $m_{\rm u}^{\rm SNII}$ $\sim$30 \ms are suggested 
by recent models accounting for metallicity dependent yields of 
SNII in full detail (\eg Timmes \etal 1995). We emphasize that the detailed 
yields for SNII at metallicities [Fe/H]$\la -1$ are not important for the 
sequential enrichment and infall model results for disk stars presented in 
this paper but we just want to note here that it is difficult to explain the 
mean [O/Fe] ratio in halo stars using the same models.

The observed breakpoint in the [O/Fe] vs. [Fe/H] relation at [Fe/H]$\sim- 1.0 
\pm$0.2 (\eg King 1994) is generally associated with the time SNIa start to 
contaminate the global disk ISM (\eg Gilmore \& Wyse 1991; Bravo \etal 1993; 
Ishimaru \& Arimoto 1995). In our models, the breakpoint in the [O/Fe] vs. 
[Fe/H] relation is mainly determined by: 1) the assumed fraction 
$\phi^{\rm SNIa}$ = 0.005 of main-sequence stars with initial masses between 
2.5 and 8 \ms (\eg Nomoto \etal 1984), 2) the delay time $\tau^{\rm SNIa}$ 
= 2.5 Gyr after which SNIa start to contribute to the enrichment of the ISM 
(\eg Smecker-Hane \& Wyse 1992; Ishimaru \& Arimoto 1995), and 3) the 
frequency distribution of SNIa as a function of age for a given stellar 
generation (assumed to be constant from $\tau^{\rm SNIa}$ to 
$\tau^{\rm SNIa} +$ 0.5 Gyr, and zero otherwise). These assumptions, in 
particular for $\tau^{\rm SNIa}$, strongly affect the increase of [Fe/H] 
with disk age and thus determine the scatter in and the slope of the 
[O/H] vs. [Fe/H] relation predicted. In fact, SNIa provide a background 
signal of iron-group elements on top of which stellar abundance variations 
due to sequential enrichment by SNII+SNIb/c occur. 

We assumed $\tau^{\rm SNIa}$ = 2.5 Gyr for the models presented in this paper,
as recently suggested by Ishimaru \& Arimoto (1995). 
However, this assumption implies different breakpoints in the [O/Fe] vs. 
[Fe/H] relation for models with disinct star formation and infall histories.
The model shown in Fig. 7-1 predicts [Fe/H]=$-1$ after $\sim$1.3 Gyr (and 
results [Fe/H]$\sim-$0.75 after 2.5 Gyr). This may indicate that the assumed 
value of $\tau^{\rm SNIa}$= 2.5 Gyr is in error. Estimates for 
$\tau^{\rm SNIa}$ based 
on stellar evolution calculations suffer from large uncertainties in the 
detailed evolution scenario for SNIa progenitors (see \eg Smecker-Hane \& 
Wyse 1992; King 1994) while theoretical estimates for $\tau^{\rm SNIa}$ based 
on the observed breakpoint may suffer from large uncertainties in the assumed 
chemical evolution of the halo (\eg Ishimaru \& Arimoto 1995).

Alternatively, model assumptions related to: 1) the enrichment rate of the 
disk ISM by SNII (\eg IMF and SFR, $m_{\rm u}^{\rm SNII}$ and SNII yields), 
or 2) the initial abundances of the material to which the SNII ejecta are 
mixed (\eg disk mass at the onset of star formation, amount of gas infall, 
and infall abundances) may be in error. For instance, adopted iron yields 
for SNII at metallicities [Fe/H]$\la -$1 may be too high by about a factor of 
two. This would be consistent with the discrepancy in the [O/Fe] ratio 
for Galactic halo stars discussed above. 
Furthermore, large amounts of metal-poor gas infall would 
improve the consistency with the assumption of $\tau^{\rm SNIa} = 2.5$ Gyr. 
Other possibilities, such as higher values of the disk total-to-final 
mass-ratio or larger SFR decay times $t_{\rm decr}$, seem to be excluded by 
the observations (\eg van den Hoek \etal 1996).

We conclude that combined sequential enrichment and metal-poor gas infall 
models are consistent with the observed plateau value and breakpoint 
in the [O/Fe] vs. [Fe/H] relation provided that \eg the adopted SNII yields at 
low metallicities [Fe/H]$\la -$1 are too high by about a factor of two.
At the same time, we have illustrated how sensitive our model results are to  
specific assumptions related to the enrichment by \eg SNII and SNIa.
These assumptions may affect quantitive conclusions concerning the 
relative importance of sequential stellar enrichment and metal-poor gas 
infall. However, our qualitative conclusion regarding the simultaneous 
presence of these processes in the local Galactic disk is not altered.

\subsection{Observational support and implications for Galactic chemical 
evolution}

We briefly discuss observational evidence in support of sequential 
stellar enrichment and metal-deficient gas infall in the local disk ISM, and 
consider possible implications for the chemical evolution of the Galaxy as a 
whole.

Sequential star formation has been argued to 
occur in nearby molecular cloud complexes including the well known
Orion, Taurus-Auriga-Perseus, Cepheus, Carina, and Chameleon cloud complexes
(\eg see the review by Blaauw 1991; Megeath \etal 1995; Testi \etal 1995; 
Goldsmith 1995).
In the Orion molecular cloud complex, age differences between OB 
subgroups are typically $\sim$2$-$7 Myr (\eg Genzel \& Stutzki 1989; 
Cunha \& Lambert 1992). 
When star formation proceeds on such time scales, massive stars 
belonging to different OB subgroups may enrich the ambient molecular cloud 
material before a next round of star formation is initiated.
Interestingly, the oldest OB subgroups in 
the Orion OB1 association appear to have oxygen abundances that are lower by 
about 40\% ($\sim$0.2 dex in [O/H]) compared to the younger subgroups 
(see \eg Olive \& Schramm 1982; Cunha \& Lambert 1992). 

Apart from observational support for sequential enrichment in nearby star 
forming regions, there are strong indications that the molecular cloud out of 
which the Sun formed has been enriched sequentially as well.
Studies related to extinct radioactive nuclides such as $^{53}$Mn 
both in the Sun and in meteorites suggest that the protosolar molecular cloud 
has been enriched by high mass stars from a 
preceding OB association, about 10$-$25 Myr prior to the actual formation of 
the Sun (\eg Cameron 1993; Swindle 1993). 

Observations suggest that high-velocity inflow of metal-deficient 
gas towards the Galactic disk is a common phenomenon as well
(\eg de Boer \& Savage 1984; Wakker 1990; Schwarz \etal 1995).
Estimates of the current gas infall rate onto the Galactic disk range 
from 0.2$-$0.5 \ms yr$^{-1}$, based on high-velocity clouds (HVCs; $v \ga$ 250 
km s$^{-1}$; \eg Mirabel \& Morras 1984; L\'{e}pine \& Duvert 1994), to 
$\sim$0.7 \ms yr$^{-1}$ derived from the soft X-ray background (Cox \& Smith 
1976), and $\sim$1.5 \ms yr$^{-1}$ based on observations of atomic hydrogen 
(Oort 1970). Since gas infall rates derived from the inflow of HVCs are likely 
underestimates (\eg Mirabel \& Morras 1984), we estimate that $\sim$30$-$40 
\% of the stars currently forming in the Galactic disk may be associated with 
infall (assuming a current SFR of $\sim 3.5$ \ms yr$^{-1}$; \eg Dopita 1987).
 
Observational support for infall induced star formation in the SNBH
has recently been presented for the Orion cloud complex (L\'{e}pine \& 
Duvert 1994; Meyer \etal 1994), the Gould Belt (\eg C\'{o}meron \& Torra 1994),
and the $\zeta$ Sculptoris open cluster (Edvardsson \etal 1995). 
These observations suggest that the most prominent star forming regions in 
the SNBH have been partly formed from infalling clouds from the Galactic halo. 
Circumstantial evidence for HVC impacts on the Galactic disk is based on: 
1) the existence of subgroups of young stars in star forming 
molecular clouds at high Galactic latitudes (like the Orion molecular cloud 
complex), 2) the displacement of OB star clusters with respect to the 
centers of their parent molecular clouds, 3) the alignment of the OB 
clusters in directions that are substantially inclined to the Galactic plane,
4) the age sequence of the aligned OB associations with an age 
of $\sim$10$^{7}$ yr for the oldest subgroups, and 5) the large elongated or 
filamentary structures observed in \eg the Orion, Taurus, Monoceros molecular 
clouds connecting the clouds to the Galactic plane (see also Tenorio-Tagle 
\etal 1986; Franco \etal 1988; G\'{o}mez de Castro 1992).
Many of these phenomena can be naturally explained by the interaction of 
a high velocity cloud with disk ISM and are difficult to reproduce by the 
process of sequential star formation (L\'{e}pine \& Duvert 1994).
Consequently, infall induced star formation appears to be a process 
frequently operating in the Galactic disk.

We have argued that metal-deficient gas infall and sequential stellar 
enrichment are likely to play an important role for the inhomogeneous 
chemical evolution of 
the Galactic disk. These processes probably occur on scales ranging from 
single molecular clouds to the formation sites of open and globular clusters
(\eg Brown 1991).
Large abundance variations of $\sim$0.3 dex in [O/Fe] observed among
metal-poor halo stars (\eg Bessell \etal 1991; Nissen \etal 1994) suggest 
that sequential enrichment and gas accretion have been important for the 
chemical evolution of the Galactic halo as well.

Nevertheless, combined metal-poor gas infall and sequential stellar 
enrichment may be a too schematic picture of the complex set of processes 
directing the chemical evolution of the Galactic disk. In particular, merger 
events with small companion galaxies may be important as well 
(\eg Quinn \etal 1993). In such case, both gas and stars in the companion 
galaxy may add substantially to the observed abundance inhomogeneities in the
Galactic disk (Pilyugin \& Edmunds 1995b). However, the suggestion that 
stellar populations from merging companion galaxies contribute substantially 
to the observed stellar abundance variations in the Galactic disk seems 
difficult to reconcile with: 1) the apparent homogeneous distribution of 
these variations within the metallicity range observed at a given age of the 
disk, and 2) the small scatter observed in the element-to-element variations 
for stars in the SNBH. Notwithstanding, merging may be important for the 
chemical evolution of the Galaxy by adding large amounts of predominantly 
metal-poor material and initiating star formation in the disk. 

Inhomogeneous chemical evolution due to sequential stellar enrichment and/or 
metal-poor gas infall is probably important also in nearby galaxies such as 
the Magellanic Clouds and M31. In the Large Magellanic Cloud, large 
abundance variations of $\sim$0.4$-$0.8 dex in [Fe/H] among similarly aged 
open clusters are observed (\eg Cohen \etal 1982; Da Costa 1991; Olsewski 
\etal 1991). Part of the variations may be accounted for by a 
radial gradient of $\sim$0.15 dex kpc$^{-1}$ in [Fe/H] (Kontizas \etal 1993).
However, the main part of these variations is likely due to triggered star 
formation in supershells as indicated by the close association of HII 
complexes with large HI holes observed in the LMC (Dopita 1985; Lortet \& 
Testor 1988). In addition, gas infall may have affected the chemical 
evolution of the tidally interacting Magellanic Clouds. 

Observational evidence in support of shock-induced star formation by SNII in 
the spiral arms of M31 has been presented by Magnier \etal (1992).
At these sites, young OB stars are observed to initiate recent star formation 
so that large abundance inhomogeneities due to sequential stellar enrichment 
are expected, similar to those observed among OB associations in the Orion star 
forming cloud complex in our own Galaxy.

Stellar and nebular abundance indicators reveal that substantial abundance 
fluctuations exist in the ISM of gas-rich galaxies (\eg Roy \& Kunth 1995). 
For instance, abundance inhomogeneities in metal-poor galaxies such as 
IZw 18 may be among the largest observed in external galaxies (\eg Kunth \etal 
1995) although this is still highly uncertain (Pettini \& Lipman 1995).
Whether the abundance fluctuations observed in dwarf galaxies are due to 
variations in self-enrichment of the H{\sc ii}-regions in these 
systems (\eg Pilyugin 1992) and/or are related to selective loss of metals 
through galactic winds driven by massive stars (Roy \& Kunth 1995; 
Martin 1996) is unclear. 

We expect that sequential stellar enrichment is generally inefficient in 
dwarf galaxies because of their low gas densities, and that the effect of 
metal-poor gas infall on the stellar abundance variations is weak due
to their low ISM abundances. Instead, star formation and abundance 
inhomogeneities induced by {\em metal-rich} gas infall may be relatively 
important in these systems.

\acknowledgements{It is a pleasure to thank L.S. Pilyugin, K. Nomoto, and 
J.-R. Roy for stimulating discussions. We like to thank J. van Paradijs for 
a critical reading of earlier versions of this paper. 
We are grateful to the referee, Dr. B. Pagel, for careful and constructive 
remarks from which this paper has benefitted.
The research of LBH is supported under grant
782-372-028 by the Netherlands Foundation for Research in Astronomy (ASTRON),
which is financially supported by the Netherlands Organisation for Scientific
Research (NWO).} \\

\

\appendix

\section{Appendix}

\noindent We describe the adopted model for the inhomogeneous 
chemical evolution of a star forming gas cloud. 
The model can be applied to various mass scales, \eg to
the entire system of molecular cloud complexes in the Galactic disk or to 
the star forming core regions within a single molecular cloud. 
We start from a homogeneous, metal free gas cloud with a 
total mass ${\rm M}_{\rm cl}$. At any evolution time $t$ in its evolution, 
this cloud is subdivided 
into ${\rm N}_{\rm scl}$ active subclouds (with corresponding masses 
M$_{\rm scl}^{i}$) involved with star formation and an inactive cloud 
part (with mass M$_{\rm qcl}$) not involved with star formation. We assume 
matter to be freely exchanged within the inactive cloud part.
Each subcloud $i$ is formed at corresponding evolution times 
$t_{\rm scl}^{i}$ and is allowed to follow its individual star formation, 
mixing, and infall history.

\subsection{Model description, definitions and assumptions}

During the lifetime $t_{\rm ev}$ of the star forming gas cloud a total number 
N$_{\rm sf}$ star formation events is assumed occur. 
Each star formation event $j$ presumably occurs within an active subcloud $i$. 
We define N$_{\rm sf}^{i}$ as the total number of star formation 
events within subcloud $i$.
For the reference model N$_{\rm sf}^{i}$ = 1 and each star formation 
event $j$ occurs in corresponding subcloud $i= j$. 
Subclouds are allowed to experience numerous star formation events, \ie 
N$_{\rm sf} > 1$. During each 
star formation event $j$ at time $t=t^{j}_{\rm sf}$ within subcloud $i$, 
a total mass of gas $\delta {\rm M}_{\rm scl}^{i} = \epsilon^{j} 
{\rm M}_{\rm scl}^{i}(t^{j}_{\rm sf})$ is transformed into stars. 

We define $\Delta t_{\rm disp}^{j}$ as the time between the 
onset of star formation within a subcloud core and the complete 
dispersal of this core region by supernova explosions and/or stellar winds.
During $\Delta t_{\rm disp}^{j}$ the subcloud core is assumed to form stars. 
The profile of the star formation 
rate (SFR) during $\Delta t_{\rm disp}^{j}$ is assumed constant and 
identical for all star formation events. However,
quantities such as the minimum stellar mass formed and IMF-slope are 
allowed to vary from one star formation event to another (\cf Sect. 4.2).
The subcloud core dispersal time determines the mass of the most massive 
star that is able to enrich subcloud cloud material before the core 
ultimately breaks up. At time of core dipsersal, the newly formed generation of 
stars has returned an amount of material 
$\delta {\rm M}_{\rm ret}^{j}$. Accordingly, the $net$ amount of material 
converted into stars during star formation event $j$ is given by: 
$\delta {\rm M}_{\rm sf}^{j} = \epsilon^{j} \delta {\rm M}_{\rm scl}^{i}
(t^{j}_{\rm sf}) - \delta {\rm M}_{\rm ret}^{j}$. 

Subclouds M$_{\rm scl}^{i}$ are formed from the inactive cloud ISM at cloud 
evolution times $t=t_{\rm scl}^{i}$. When a subcloud forms 
it adopts the abundances of the inactive cloud ISM at $t=t_{\rm scl}^{i}$.
For each subcloud, we define a mixing time scale 
$\Delta t_{\rm mix}^{i}$ as the time between formation of the 
subcloud and the actual break up of the entire subcloud. 
The instant of break up of the subcloud may be either after one or more star 
formation events, or before star formation actually takes place. 
In this manner, material can be deposited 
within a subcloud region for a considerable period of time
before being mixed to the surrounding ISM.
The mixing history of each subcloud directs both the inhomogeneous chemical 
evolution of the inactive cloud and that of the neighboring subclouds.

Before an entire subcloud breaks up its constituent material will be enriched 
by the stellar populations it is hosting. We assume the stellar enrichment 
of the subcloud to proceed homogeneously. In order to allow for sequential 
enrichment, we consider a fraction $\lambda^{j}$ of
enriched material ejected {\em during} star formation event $j$ to mix 
homogeneously 
with subcloud core material hosting the {\em next} star formation event. 
Simultaneous with the ejection of enriched material returned by newly 
formed stars, a substantial fraction of the ambient subcloud matter 
$\kappa^{j} {\rm M}_{\rm scl}^{i}$ may be swept up during dispersal of 
its star forming core. This subcloud material may mix to the 
subcloud hosting the next star formation event as well.
The subcloud hosting the next star formation event may be either 
the subcloud hosting the current star formation event or a subcloud nearby. 
No matter exchange is assumed between the subcloud and the surrounding ISM
during the time between two star formation events occuring within one and the
same subcloud. In case of the reference model, we do not consider 
mass transfer between subclouds, \ie $\lambda^{j} = \kappa^{j} = 0$.

After an entire subcloud breaks up its matter is assumed to mix homogeneously 
to the inactive cloud part. At the same time, stars associated with 
the dispersing subcloud become part of the stellar populations in
the inactive cloud. 
After break up, different cloud fragments present in the ambient ISM may form 
new subclouds wherein star formation occurs as soon as the critical conditions 
for star formation are met.

In addition to the individual chemical evolution of subclouds, which is 
directed by 
their star formation history and exchange history with the surrounding 
ISM, we allow for local enrichment of a given subcloud by stars that 
were {\em not} formed within that subcloud. 
This may be particularly important for low mass 
SNIa-progenitors 
which travelled considerable distances from their birth sites
and enrich their immediate surroundings at the time they explode as SNIa
(\cf Fig. 2f; see below).

\subsection{Basic equations}

We keep track of the total mass of and abundances in stars and gas as a 
function of evolution time, both within each subcloud and the inactive cloud 
ISM. For each star 
formation event we use conventional chemical evolution model equations (\eg 
Tinsley 1980, see below) except for including metallicity 
dependent stellar lifetimes, remnant masses and element yields 
(\cf van den Hoek \etal 1996). 

\subsubsection{Mass-exchange between subclouds and the inactive cloud ISM}

We denote $\Delta Q$ as the variation of a quantity $Q$ between two
cloud evolution times $t - \Delta t$ and $t$. 
With M$_{\rm cl}(t=0)$  the initial mass of the 
cloud and no stars initially present,
\ie M$_{*}(0)$ = 0, we can express the variations of mass of gas and stars 
within the cloud as:
\begin{eqnarray}
\Delta {\rm M}_{\rm cl} & = & \Sigma_{i=1}^{N_{\rm scl}(t)} 
        \Delta {\rm M}_{\rm scl}^{i} + \Delta {\rm M}_{\rm qcl} \\ 
\Delta {\rm M}_{\rm *} & = & \Sigma_{j=1}^{N_{\rm sf}(t)} 
\left( \Delta {\rm C}_{*}^{j} - \Delta {\rm E}_{*}^{j} \right) 
        - \Delta {\rm E}_{*, {\rm qcl}}
\end{eqnarray}
where N$_{\rm scl}(t)$ is the current number of individual subclouds,
N$_{\rm sf}(t)$ the current number of star formation events 
within the cloud, $\Delta {\rm C}^{j}$ the total mass of stars 
formed during star formation event $j$, and $\Delta {\rm E}^{j}$ the total 
mass of matter returned within $\Delta t$ by stars formed during star 
formation event $j$. We recall conventional expressions for 
$\Delta {\rm C}^{j}$ and $\Delta {\rm E}^{j}$ (\cf Tinsley 1980):
\begin{eqnarray}
& & \Delta {\rm C}^{j} = 
\int_{t_{\rm sf}}^{t_{\rm sf}+t_{\rm disp}}
\int_{m_{\rm l}}^{m_{\rm u}} m {\rm S}_{j}(t) {\rm M}_{j}(m) \: \: 
{\rm d}m {\rm d}t \\
& & \Delta {\rm E}^{j} = \nonumber \\
& & \int_{t-\Delta t}^{t} 
\int_{m_{\rm o}(t-t_{\rm sf})}^{m_{\rm o}(t-
t_{\rm sf}-t_{\rm disp})} 
\!\!\!\!\!\!\!\!\!\!\!\!\!\!\!\!\!\!\!\!\!\!\!\!\!\!\!
\left( m-m_{\rm rem}(m) \right)
{\rm S}_{j}(t- \tau(m)) {\rm M}_{j}(m) \:\: {\rm d}m {\rm d}t
\end{eqnarray}
where S$_{j}$ and M$_{j}$ denote the SFR by number [yr$^{-1}]$ and
IMF [\mss$^{-1}]$ for star formation event $j$. For convenience, we ignored the 
index $j$ for 
$t_{\rm sf}$, $t_{\rm disp}$ as well as for the stellar mass boundaries at 
birth $m_{\rm l}$, $m_{\rm u}$.
We emphasize that both the stellar remnant masses $m_{\rm rem}(m)$, 
lifetimes $\tau(m)$, and turnoff-masses $m_{\rm o}(t)$ are a 
function of the initial metallicity $Z_{*}$ (containing all elements 
heavier than He) of the stellar generation under 
consideration. Variations in the total gas masses within the inactive cloud and 
subcloud $i$, \ie M$_{\rm qcl}$ and M$_{\rm scl}^{i}$ respectively, can be 
expressed as:
\begin{eqnarray}
\Delta {\rm M}_{\rm qcl} & = & \Delta {\rm E}_{*, {\rm qcl}} -
\Sigma_{\rm form} {\rm M}_{\rm scl}^{k} + \Sigma_{\rm disp} 
{\rm M}_{\rm scl}^{l} \\
\Delta {\rm M}_{\rm scl}^{i} & = &
\left[ \Delta {\rm E}_{*} - \Delta {\rm C}_{*} \right]^{i} + 
\Delta {\rm M}_{\rm sf, prev} - \Delta {\rm M}_{\rm sf, next} \\ 
\Delta {\rm M}_{\rm scl} & = & \Sigma_{i=1}^{{\rm N}_{\rm scl}(t)}
\Delta {\rm M}_{\rm scl}^{i}  + \Sigma_{\rm form} {\rm M}_{\rm scl}^{k} 
- \Sigma_{\rm disp} {\rm M}_{\rm scl}^{l} 
\end{eqnarray}
where $\Delta {\rm E}_{\rm *, qcl}$ refers to the amount of material returned 
by stars present in the inactive cloud within time $\Delta t$. 
We followed both the stellar ejecta from recently formed stars
within active subclouds and the ejecta from older stellar populations 
present in the inactive cloud ISM.
The total amount of gas depleted by subclouds which are formed 
within time $\Delta t$ is denoted by $\Sigma_{\rm form} 
{\rm M}_{\rm scl}^{k}$. Similarly, the amount of gas returned by subclouds 
which become dispersed within time $\Delta t$ is denoted by 
$\Sigma_{\rm disp} {\rm M}_{\rm scl}^{l}$. 
We remark that the term between square brackets in Eq. (A6) refers to star 
formation events which occur {\em within} subcloud $i$.

\subsubsection{Supernovae Type Ia}

The term $\Delta {\rm E}_{*}^{i}$ in Eq. (A6) is related both to stellar 
generations which formed within subcloud $i$ 
and to stars that entered the subcloud from elsewhere
in the cloud.
We will consider the case of SNIa progenitors stars only. Consequently, 
the term $\Delta {\rm E}_{*}^{i}$ can be expressed as two terms, \ie 
$\Delta {\rm E}_{*}^{i} = \Sigma_{\rm sf} 
( \Delta {\rm E}^{j})_{\rm scl}^{i} + \Delta {\rm E}_{\rm SNIa}$. The 
former term is related to star formation events which occured within subcloud 
$i$ while the latter term is associated with subcloud enrichment by 
SNIa-progenitors formed elsewhere in the cloud. 
We define the total amount of matter returned by SNIa within subcloud $i$ 
during time $\Delta t$ as: $\Delta {\em E}_{\rm SNIa} \equiv 
\alpha_{\rm SNIa}^{i} \Delta t R_{\rm SNIa} m_{\rm rem}(m)$ where 
$R_{\rm SNIa}$ is the total average SNIa-rate in the entire cloud and 
$\alpha_{\rm SNIa}^{i}$ the corresponding fraction of SNIa that is
is assumed to go off within subcloud $i$. In case of the reference model 
$\Delta {\rm E}_{\rm SNIa}$ =0.

\subsubsection{Mass-exchange between individual subclouds}

As matter may be transferred from one subcloud to another (or within one 
subcloud from one subcloud core to another) we include terms 
$\Delta {\rm M}_{\rm sf, prev}$ and $\Delta {\rm M}_{\rm sf, next}$ in 
Eq. (A6). The term $\Delta {\rm M}_{\rm sf, prev}$ corresponds to the amount 
of material added from the {\em preceding} star formation event to the core of
the subcloud currently experiencing star formation. The term 
$\Delta {\rm M}_{\rm sf, next}$ refers to the 
amount of matter mixed from the subcloud core actually experiencing star 
formation to the subcloud core hosting the {\em next} star formation event.
For each star formation event $j$ which happens to occur in subcloud $i$ 
within the time interval $\Delta t$ we may write:
\begin{eqnarray}
\Delta {\rm M}_{\rm sf, prev} & = & \lambda^{j-1} 
\delta {\rm M}_{\rm ret}^{j-1} + \kappa^{j-1} {\rm M}_{\rm scl, prev} \\
\Delta {\rm M}_{\rm sf, next} & = & \lambda^{j} 
\delta {\rm M}_{\rm ret}^{j} + \kappa^{j} {\rm M}_{\rm scl}^{i}
\end{eqnarray}
where $M_{\rm scl, prev}$ is the mass of the subcloud hosting the preceding 
star formation event. In this paper, we presented only results for 
$\kappa^{j} = 0$. In general, $\kappa^{j}$ $>$0 has a similar effect  
as when reducing the sequential enrichment efficiency 
$\lambda^{j}$.

\subsubsection{Chemical evolution of subclouds and inactive cloud ISM}

Expressions for the average abundance changes of element $X$ within the 
entire cloud, inactive cloud part, and subclouds can be written as:
\begin{eqnarray}
& & \Delta (X_{\rm cl} {\rm M}_{\rm cl}) =
\Sigma_{i=1}^{{\rm N}_{\rm scl}(t)}
\Delta (X_{\rm scl}^{i} {\rm M}_{\rm scl}^{i}) + \Delta (X_{\rm qcl} 
{\rm M}_{\rm qcl}) \\
& & \Delta (X_{\rm qcl} {\rm M}_{\rm qcl}) = \Delta {\rm E}_{X {\rm , qcl}}
- \Sigma_{\rm form} X_{\rm qcl} {\rm M}_{\rm scl}^{k} +
\Sigma_{\rm disp} X_{\rm scl}^{l} {\rm M}_{\rm scl}^{l} \\
& & \Delta (X_{\rm scl} {\rm M}_{\rm scl}) = \nonumber \\
& & \Sigma_{i=1}^{{\rm N}_{\rm scl}(t)}
\Delta (X_{\rm scl}^{i} {\rm M}_{\rm scl}^{i}) 
+ \Sigma_{\rm form} X_{\rm qcl} {\rm M}_{\rm scl}^{k} -
\Sigma_{\rm disp} X_{\rm sql}^{l} {\rm M}_{\rm scl}^{l} \\
& & \Delta (X_{\rm scl}^{i} {\rm M}_{\rm cl}^{i}) = \left[ 
\Delta {\rm E}_{X} -X_{\rm scl} \Delta {\rm C}_{*}  \right]^{i} 
\Delta {\rm M}_{X {\rm , pre}} - \Delta {\rm M}_{X {\rm , next}}
\end{eqnarray}
where the meaning of each term can be found from its counter 
part in Eqs. A5-A7.
Similarly, expressions for $\Delta {\rm M}_{X {\rm , prev}}$
and $\Delta {\rm M}_{X {\rm , next}}$ can be written as:
\begin{eqnarray}
\Delta {\rm M}_{X {\rm , prev}} & = & \lambda^{j-1}
\delta {\rm M}_{X {\rm , ret}}^{j-1}
+ \kappa^{j-1} (X_{\rm scl} {\rm M}_{\rm scl})_{\rm prev} \\
\Delta {\rm M}_{X {\rm , next}} & = & \lambda^{j}
\delta {\rm M}_{X {\rm , ret}}^{j}
+ \kappa^{j} X_{\rm scl}^{i} {\rm M}_{\rm scl}^{i} 
\end{eqnarray}
where $\Delta {\rm E}_{X}^{j}$ is the total mass of enriched material of 
element $X$ returned within time $\Delta t$ by a stellar generation formed 
during star formation event $j$:
\begin{eqnarray}
& & \Delta {\rm E}_{X}^{j} = \nonumber \\
& & \int_{t- \Delta t}^{t} 
\int_{m_{\rm o}(t-t_{\rm sf})}^{m_{\rm o}(t-
t_{\rm sf}-t_{\rm disp})} 
\!\!\!\!\!\!\!\!\!\!\!\!\!\!\!\!\!\!\!\!\!\!\!\!\!\!\!
\Delta {\rm M}_{X}(m) 
{\rm S}_{j}(t- \tau(m)) {\rm M}_{j}(m) \:\: {\rm d}m {\rm d}t \\
& & \Delta {\rm M}_{X}(m) = mp_{\rm X}(m) + \left( m- m_{\rm rem}(m)
\right) X_{*}^{j}
\end{eqnarray}
and $\Delta {\rm M}_{X}(m)$ is the total mass of element $X$ ejected by a 
star of initial mass $m$ born with metallicity $X_{*}^{j}$ during star 
formation event $j$. The term $\Delta {\rm M}_{X}(m)$ includes both
newly synthesized stellar material and matter initially present at the time 
stars were formed. Initial stellar abundances $X_{*}^{j}$ are determined 
by the abundances of the subcloud $i$ (hosting star formation event $j$) at time
$t_{\rm sf}^{j}$, \ie $X_{*}^{j} = X_{\rm scl}^{i} (t=t_{\rm sf}^{j})$.
Literature sources for the adopted theoretical metallicity dependent stellar 
yields $p_{\rm X}(m)$, stellar lifetimes $\tau (m)$, and remnant masses 
$m_{\rm rem}(m)$ are given in Sect. 3.4. \\

\noindent {\bf References}

\begin{footnotesize}
\begin{list}{}{\reflistset}
\item
Anders E. \& Grevesse N. 1989, Geochim. Cosmochim. Acta 53, 197
\item
Andersson H. \& Edvarsson B. 1995, A\&A 290, 590
\item
Baldwin J.A., Ferland G., Martin P.G., \etal 1991, ApJ 374, 580
\item
Bahcall J.N. \& Pinsonneault M.H. 1995, to appear in: Review of Modern 
 Physics
\item
Bahcall J.N. \& Soneira R.M. 1980, ApJS 44, 73
\item
Basu S. \& Rana N.C. 1992, Ap\&SS 196, 1
\item
Bateman N.P.T. \& Larson R.B. 1993, ApJ 407, 634 
\item
van den Bergh S. \& Tammann G.A. 1991, ARA\&A 29, p.\ 363
\item
Bessell M.S., Sutherland R.S., Ruan K. 1991, ApJ 383, L71 
\item
Binney J. \& Tremaine S. 1987, Galactic Dynamics, Princeton Univ. Press, 
  Princeton
\item
de Boer K.S. \& Savage B.D. 1984, A\&A 136, L7
\item
Blaauw  A. 1991, in: Lada J., Kylafis N.D. (eds.) The Physics of Star 
  Formation and Early Stellar Evolution, NATO ASI Series
\item
Boesgaard A.M. 1989, ApJ 336, 798
\item
Bravo E., Isern J., Canal R. 1993, A\&A 270, 288
\item
Brown J.H. 1991, Ph.D. Thesis, Univ. of Illinois, Urbana-Champaign. 
\item
Buonanno R., Corsi C.E., Fusi Pecci F. 1989, A\&A 216, 80
\item
Cameron A.G.W. 1993, in: Levy E.H., Lunine J.I (eds.)
  Protostars and Planets III, Univ. of Arizona Press, p.\ 43
\item
Capellaro E., Turatto H, Benetti S., \etal. 1993, A\&A 273, 383
\item
Carlberg R.G., Dawson P.C., Hsu T., van den Berg D.A. 1985, ApJ 294, 674 
\item
Carney B.W., Latham D.W., Laird J.B. 1990, AJ 99, 752
\item
Carraro G. \& Chiosi C. 1994, A\&A 287, 761
\item
Cioffi D.F. \& Shull J.M. 1991, ApJ 367, 96
\item
Clayton D.D. 1988, MNRAS 234, 1
\item
Cohen J.G. 1982, ApJ 258, 143
\item
C\'{o}meron F. \& Torra J. 1994, A\&A 281, 35
\item
Cox D.P \& Smith B.W. 1974, ApJ 189, L105
\item
Cunha K., \& Lambert D.L. 1992, ApJ 399, 586
\item
Da Costa G.S. 1991, in: Haynes R.F., Milne D.G. (eds.) The Magellanic Clouds 
  and their Dynamical Interaction with the Milky Way, IAU Symp. 148, Kluwer, 
  Dordrecht, p.\ 183
\item
Dopita M.A. 1985, ApJ 295, L5
\item
Dopita M.A. 1987, in: Faber S.M. (ed.) Nearly Normal Galaxies from the 
  Planck Time to Present, Springer, New York, p.\ 144
\item
Dopita M.A. 1990, in: Thronson H.A., Shull J.M. (eds.) The Interstellar 
  Medium in Galaxies, Kluwer, Dordrecht, p.\ 437
\item
Edmunds M.G. 1975, Ap\&SS 32, 483
\item
Edmunds M.G., 1993 Nat 365, 293
\item
Edvardsson B., Andersen J., Gustafsson B., \etal 1993a, A\&A 275, 101 
({\bf EDV})
\item
Edvardsson B., Andersen J., Gustafsson B., \etal 1993b, A\&AS 102, 603
\item
Edvardsson B., Petterson B., Kharrazi M., Westerlund B. 1995, A\&A 293, 75
\item
Elmegreen B.G. \& Lada C.J. 1977, ApJ 214, 725
\item
Fich M. \& Tremaine S. 1991, ARA\&A 29, 409
\item
Fitzsimmons A., Brown P.J.F., Dufton P.L., and Lennon D.J. 1990, A\&A 250, 159
de Freitas Pacheco J.A. 1993, ApJ 403, 673
\item
Franco J., Tenorio-Tagle G., Bodenheimer P., R\'{o}zyczka M., Mirabel I.F., 
 1988, ApJ 333, 826
\item
Francois P., \& Matteucci F. 1993, 280, 136
\item
Friel E.D. \& Janes K.A. 1993, A\&A 267, 75
\item
Garci\'{a}-Lopez \etal 1993, ApJ 412, 173
\item
Garmany C.D., Conti P.S., Chiosi C., 1982, ApJ 263, 777
\item
Gehren T., Nissen P.E., Kudritzki R.P., Butler K. 1985, in: Danziger I.J, 
 Matteuci F., Kjaer K. (eds.) Abundance Gradients in the Galactic Disk from 
 young B-type Stars in Clusters, ESO Conf. and Workshop Proc. 21, p.\ 171
\item
Genzel R. \& Stutzki J. 1989, ARA\&A 27, 41
\item
Gies D.R., and Lambert D.L. 1992, ApJ 387, 673
\item
Gilmore G. 1989, in:  Buser R., King I.R. (eds.) The Milky Way as a Galaxy, 
Univ. Science Books, Mill Valley Ca., p.\ 281
\item
Gilmore G., \& Wyse R.F.G. 1991, ApJ 367, L55
\item
Goldsmith P.F. 1995, to appear in: Chiao R.Y. (ed.) C.H. Townes Festschrift,
  Arecibo preprint
\item
G\'{o}mez de Castro A.I. 1992, in: Palous J., Burton W.B., Lindblad P.O 
  (eds.) Evolution of the Interstellar Medium and Dynamics of Galaxies, 
  Cambridge Univ. Press
\item
Gratton R.G. \& Sneden C. 1991, A\&A 241, 501
\item
Grenon M. 1987, JA\&A 8, 3
\item
Grenon M. 1989, Ap\&SS 156, 29
\item
Grevesse N. \& Noels A. 1993, Phys. Scripta T47, 133
\item
Groenewegen M.A.T. \& de Jong T. 1993, A\&A 267, 410
\item
Groenewegen M.A.T., van den Hoek L.B., de Jong T., 1995, A\&A 293, 381
\item
Haikala L.K. 1995, A\&A 294, 89
\item
Hashimoto M., Iwamoto K., Nomoto K. 1993, ApJ 414, L105
\item
Hashimoto M., Nomoto K., Tsujimoto T., Thielemann F.-K. 1993, in: 
  K\"{a}ppeler F., Wisshak K. (eds.) Nuclei in Cosmos, Institute Physics Publ., 
  p.\ 587
\item
Henning T. \& G\"{u}rtler J. 1986, Ap\&SS 128, 199
\item
van den Hoek L.B., Groenewegen M.A.T., Nomoto K., and de Jong T. 1996, A\&A
{\em in preparation}
\item
Ishimaru Y. \& Arimoto N. 1995, A\&A, {\em submitted}
\item
Kaufer A., Szeifert Th., Krenzin R., Baschek B., Wolf B. 1994, A\&A 289, 740
\item
Kennicutt R.C., Jr. 1983, ApJ 272, 54
\item
King J.R. 1994, AJ 107, 350 
\item
King D.L., Vladilo G., Lipman K., \etal 1995, to appear in A\&A
\item
Klochkova V.G., Mishenina T.V. \& Panchuk V.E. 1989, Soviet Astron. Lett.
15, 135
\item
Kontizas M., Kontizas E., Michalitsianos A.G. 1993, A\&A 269, 107
\item
Kulkarni S.R.  \& Heiles C. 1987, in: Hollenbach D.J., Thronson H.A. (eds.)
 Interstellar Processes, Reidel, Dordrecht, p.\ 87
\item
Kunth D., Lequeux J., Sargent W.L.W., Viallefond F. 1994, A\&A 282, 709
\item
Kunth D., Matteucci F., Marconi G. 1995, A\&A 297, 634
\item
Lambert D.L. 1989, in: Waddington C.J. (ed.) Cosmic Abundances of Matter, 
  Amer. Inst. Phys., New York, p.\ 168
\item
Larson R.B. 1969, MNRAS 145, 504
\item
Larson R.B. 1976, MNRAS 176, 31
\item
Larson R.B., Tinsley B.M., Caldwell C.N. 1980, AJ 237, 692
\item
Lennon D.J., Dufton P.L., Fitzsimmons A., Gehren T., Nissen P.E. 1990, A\&A 
  240, 349
\item
Leisawitz D.T. 1985, Ph.D. Thesis, Univ. of Texas, Ausin
\item
L\'{e}pine J.R.D. \& Duvert G. 1994, A\&A 286, 60
\item
Lortet M.-C. \& Testor G. 1988, A\&A 194, 11 
\item
Maeder A. 1992, A\&A 264, 105
\item
Maeder A. 1993, A\&A 268, 833
\item
Magnier E.A., Lewin W.H.G., van Paradijs J., \etal 1992, A\&AS 96, 379
\item
Martin C.L., 1996, ApJ, in press
\item
Mayor M. 1976, A\&A 48, 301
\item
Mayor M. \& Martinet L. 1977, A\&A 55, 221
\item
Megeath S.T., Cox P., Bronfman L., Roelfsema P.R. 1995, to appear in A\&A
\item
Meusinger H., Reimann H.-G., and Stecklum B. 1991, A\&A 245, 57
\item
Meyer D.M., Jura M., Hawkins I., Cardelli J.A. 1994, ApJ 437, L59
\item
Mezger P.G. 1988, in: Pudritz R.E., Fich M. (eds.) Galactic and 
 Extragalactic Star Formation, Kluwer, Dordrecht, p.\ 227
\item
Mirabel I.F. \& Morras R. 1984, ApJ 279, 86
\item
Nissen P.E. 1988, A\&A 199, 46
\item
Nissen P.E., Gustafsson B., Edvardsson B., Gilmore G., 1994, A\&A 285, 440
\item
Nomoto K., Thielemann F.-K., Yokoi K. 1984, ApJ 286, 644
\item
Nomoto K., Kumugai S., Shigeyama T. 1991, in: Durouchoux P., Prantzos P.
(eds.) Gamma-Ray Line Astrophysics, AIP Conf. Proc. 232, AIP, New York, 
  p.\ 236
\item
Nomoto K. \etal 1993, Nat 364, 507
\item
Nomoto K., Iwamoto K., Tsujimoto T., Hashimoto M. 1994, in: Suzuki Y., 
  Nakamura K. (eds.) Frontiers of Neutrino 
\item
Olive K.A., \& Schramm D.N. 1982, ApJ 257, 276
\item
Olsewski E.W., Schommer R.A., Suntzeff N.B., Harris H.C. 1991, AJ 101, 515
\item
Ogelman H.B., \& Maran S.P. 1976, ApJ 209, 124
\item
Osterbrock D.E., Tran H.D., Veilleux S. 1992, ApJ 389, 305
\item
Pagel B.E.J. \& Tautvai\v{s}ien\.{e} G. 1995, MNRAS 276, 505
\item
Peimbert M. 1987, in: Star Forming Regions, Conf. Proc., Tokyo, Reidel, 
   Dordrecht, p.\ 111
\item
Peimbert M., Storey P.J., Torres-Peimbert S. 1993, ApJ 414, 626
\item
Pettini M. \& Lipman K., 1995, A\&A, in press
\item
Pismis 1990, A\&A 234, 443
\item
Pilyugin L.S. 1992, A\&A 260, 58
\item
Pilyugin L.S. \& Edmunds M.G. 1995a, A\&A, {\em submitted}
\item
Pilyugin L.S. \& Edmunds M.G. 1995b, in: 'The Interplay between Massive  
  Star Formation, the ISM and Galaxy Evolution', 11th IAP Ap. Meeting, Paris,
  Kunth D., Guiderdoni B., Heydari-Malayeri M., Thuan T.X., and 
  V\^{a}n J.T.T. (Eds.)
\item
Prantzos N. 1994, A\&A 284, 477
\item
Quinn P.J., Hernquist L., Fullagar D.P. 1993, ApJ 403, 74
\item
Rizzo J.R. \& Bajaja E. 1994, A\&A 289, 922
\item
Rolleston W.R.J., Dufton P.L., Fitzsimmons A. 1994, A\&A 284, 72
\item
Roy J.-R., \& Kunth D. 1995, A\&A 294, 432
\item
Schaller G., Schaerer D., Meynet G., Maeder A. 1992, A\&AS 96, 269
\item
Schuster W.J, \& Nissen P.E. 1989, A\&A 222, 69
\item
Schwarz U.J., Wakker B.P., van Woerden H. 1995, A\&A 302, 364
\item
Shaver P.A., McGee R.X., Newton L.M., \etal 1983, MNRAS 204, 53
\item
Shigeyama T. \etal 1990, ApJ 361, L23 
\item
Smecker-Hane T.A. \& Wyse R.F.G. 1992, AJ 103, 1621
\item
Steigman G. 1993, ApJ 413, L73
\item
Strobel A. 1991, A\&A 247, 35
\item
Strom R.G. 1993, in NATO ASI Series Vol. 450 on {\em The lives of Neutron 
 Stars}, Eds. M.A. Alpar, \"{U}. Kiziloglu, and J. van Paradijs, 23
\item
Swindle T.D. 1993, in: Levy E.H., Lunine J.I (eds.)
  Protostars and Planets III, Univ. of Arizona Press, p.\ 867
\item
Tenorio-Tagle G., Bodenheimer P., R\'{o}zyczka M., Franco J. 1986, A\&A 170, 
  107
\item
Testi L., Olmi L., Hunt L., \etal 1995, to appear in A\&A
\item
Thielemann F.-K., Nomoto K., Hashimoto M. 1994, in: Audouze J., Bludman S., 
 Mochkovitz R., Zinn-Justin J. (eds.) Supernovae, 
  Proc. Session LIV, Les Houches, Elsevier Sci. Publ. Amsterdam, p.\ 629
\item
Thuan T.X., \etal 1995, A\&A, {\em in preparation}
\item
Timmes F.X., Woosley S.E., Weaver T.A. 1995, ApJS 98, 617
\item
Tinsley B.M. 1980, Fund. Cosmic Phys. 5, 287
\item
Trimble V. 1991, A\&AR 3, 1
\item
Tsujimoto T., Nomoto K., Hashimoto M., Thielemann F.-K. 1995, submitted to 
  ApJ
\item
Tutukov V.A., Yungelson L.R., Iben I., Jr. 1992, ApJ 386, 197
\item
Twarog B.A. 1980a, ApJ 242, 242
\item
Twarog B.A. 1980b, ApJS 44, 1
\item
Twarog B.A. \& Wheeler J.C. 1982, ApJ 261, 636
\item
Wakker B.P. 1990, Ph.D. Thesis, Univ. of Groningen
\item
Walterbos R.A.M. 1988, in: Pudritz R.E., Fich M. (eds.) Galactic and 
 Extragalactic Star Formation, NATI/ASI Conf. Proc. 232, p.\ 361
\item
Wielen R., Fuchs B., Dettbarn C., 1996, A\&A, in press
\item
Wilking B.A. \& Lada C.J. 1983, ApJ 274, 571
\item
Wilmes M. \& K\"{o}ppen J. 1995, A\&A 294, 47 
\item
Wilson T.L. \& Matteucci F. 1992, A\&AR 4, 141 
\item
Woosley S.E., Langer N., Weaver T.A. 1993, ApJ 411, 823
\item
Woosley S.E., Langer N., Weaver T.A. 1995, ApJ 448, 315
\item
Yamaoka H. 1993, Ph.D. Thesis, Univ. of Tokyo
\item
York D.G., Spitzer L., Bohlin R.C., \etal 1983, ApJ 266, L55
\end{list}
\end{footnotesize}

\end{document}